%% file: PREarxiv.tex
\documentclass[reprint,superscriptaddress,twocolumn,pre]{revtex4-1}
\usepackage{times}
\usepackage{amssymb,amsmath,amsthm,amsfonts,wrapfig,esint,afterpage,marginnote,afterpage,lipsum,color,tabularx}
\usepackage[normalem]{ulem}
\usepackage{natbib}
\usepackage{lineno,url}\usepackage{graphicx,epsfig}\usepackage{xr}
\usepackage{pstricks}
\usepackage{lipsum, babel}
\usepackage{blindtext}
\input{./hudef.tex}

\input{./p2pdef.tex}
\newlength{\tew}

\def\medskip{}\def\bigskip{}

\usepackage{tikz}
 \def\Uold{U_{{\rm old}}}

\usepackage{hyperref}
 \newcommand{\Bra}[1]{\left[#1\right]}

\begin{document}
	\title{Origin of Jumping Oscillons in an Excitable Reaction-Diffusion System}
	
	
	\author{Edgar Knobloch}
	\affiliation{Department of Physics, University of California at Berkeley, Berkeley, CA 94720, USA}
	\email{knobloch@berkeley.edu}
	\author{Hannes Uecker}
	\affiliation{Institut f\"ur Mathematik, Universit\"at Oldenburg, D26111 Oldenburg, Germany}
	\email{hannes.uecker@uni-oldenburg.de}
	\author{Arik Yochelis}
	\affiliation{Department of Solar Energy and Environmental Physics, Blaustein Institutes for Desert Research, Ben-Gurion University of the Negev, Sede Boqer Campus, Midreshet Ben-Gurion 8499000, Israel}
	\affiliation{Department of Physics, Ben-Gurion University of the Negev, Be'er Sheva 8410501, Israel}
	\email{yochelis@bgu.ac.il}
	
	\date{\today}
	
	\begin{abstract}
		\noindent 
		Oscillons, i.e., immobile spatially localized but temporally oscillating structures, are the subject of intense study since their discovery in Faraday wave experiments. However, oscillons can also disappear and reappear at a shifted spatial location, becoming jumping oscillons (JOs). We explain here the origin of this behavior in a three-variable reaction-diffusion system via numerical continuation and bifurcation theory, and show that JOs are created via a modulational instability of excitable traveling pulses (TPs). We also reveal the presence of bound states of JOs and TPs and patches of such states (including jumping periodic patterns) and determine their stability. This rich multiplicity of spatiotemporal states lends itself to information and storage handling.
	\end{abstract}
	\maketitle
	
	
	Time-dependent spatially localized states in dissipative systems, such as action potentials in
	physiology~\cite{izhikevich2007dynamical,murray2007mathematical,ksp08,allard2013traveling} and oscillons in Faraday waves~\cite{umbanhowar1996localized,lioubashevski1999oscillons,blair2000patterns}, have attracted considerable interest in the past half-century~\cite{ankiewicz2008dissipative,knobloch2015spatial}. Work in these fields not only provided insights into the original model (e.g., neural and cardiac) systems but also stimulated understanding of pattern formation mechanisms, which in turn proved applicable in other settings, such as in chemical reactions~\cite{VE07}, nonlinear optics~\cite{firth2007homoclinic,barbay2008homoclinic,firth2007snaking},
	fluid convection~\cite{jacono2011magnetohydrodynamic,alonso2011convectons,mercader2011convectons,watanabe2012spontaneous}, and even sound discrimination in the inner ear~\cite{edri2018molding}. Importantly, advances in pattern formation theory require the development of nonlinear methodologies and, specifically, numerical (continuation) methods for differential equations~\cite{knobloch2015spatial}, such as the packages AUTO~\cite{doedel2012auto}, MatCont~\cite{bindel14}, and pde2path~\cite{p2pbook}. These assist in revealing complex pattern selection mechanisms via computation of both stable and unstable solutions {upon variation of control parameters}, and so provide insights that are impossible to uncover otherwise. Thus, continuation methods are essential means of analysis for spatially extended complex systems regardless of the nature of the model equations.
	
	\begin{figure}[bp]
		\centering
		(a)\includegraphics[width=0.21\textwidth]{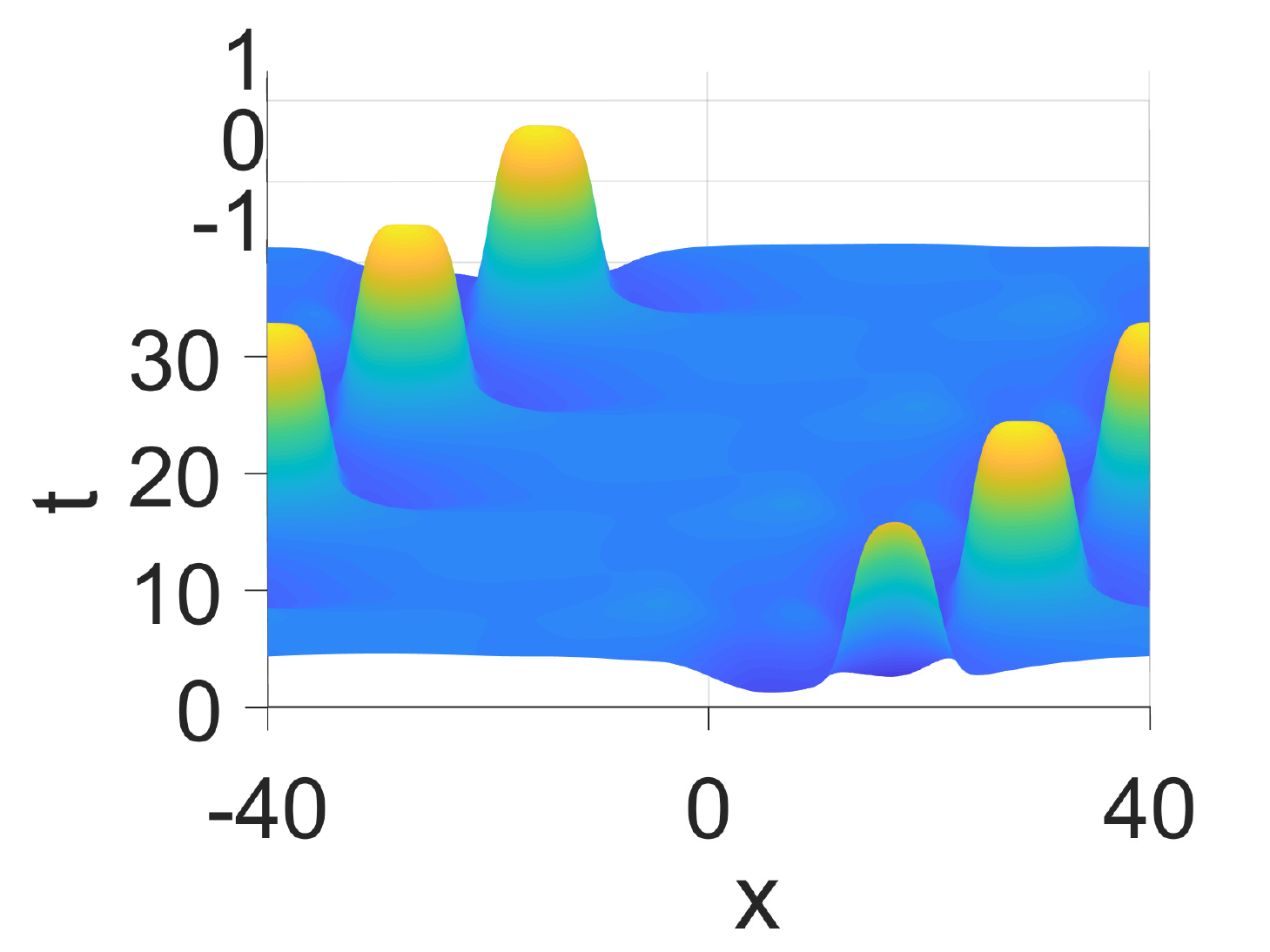}
		(d)\includegraphics[width=0.21\textwidth]{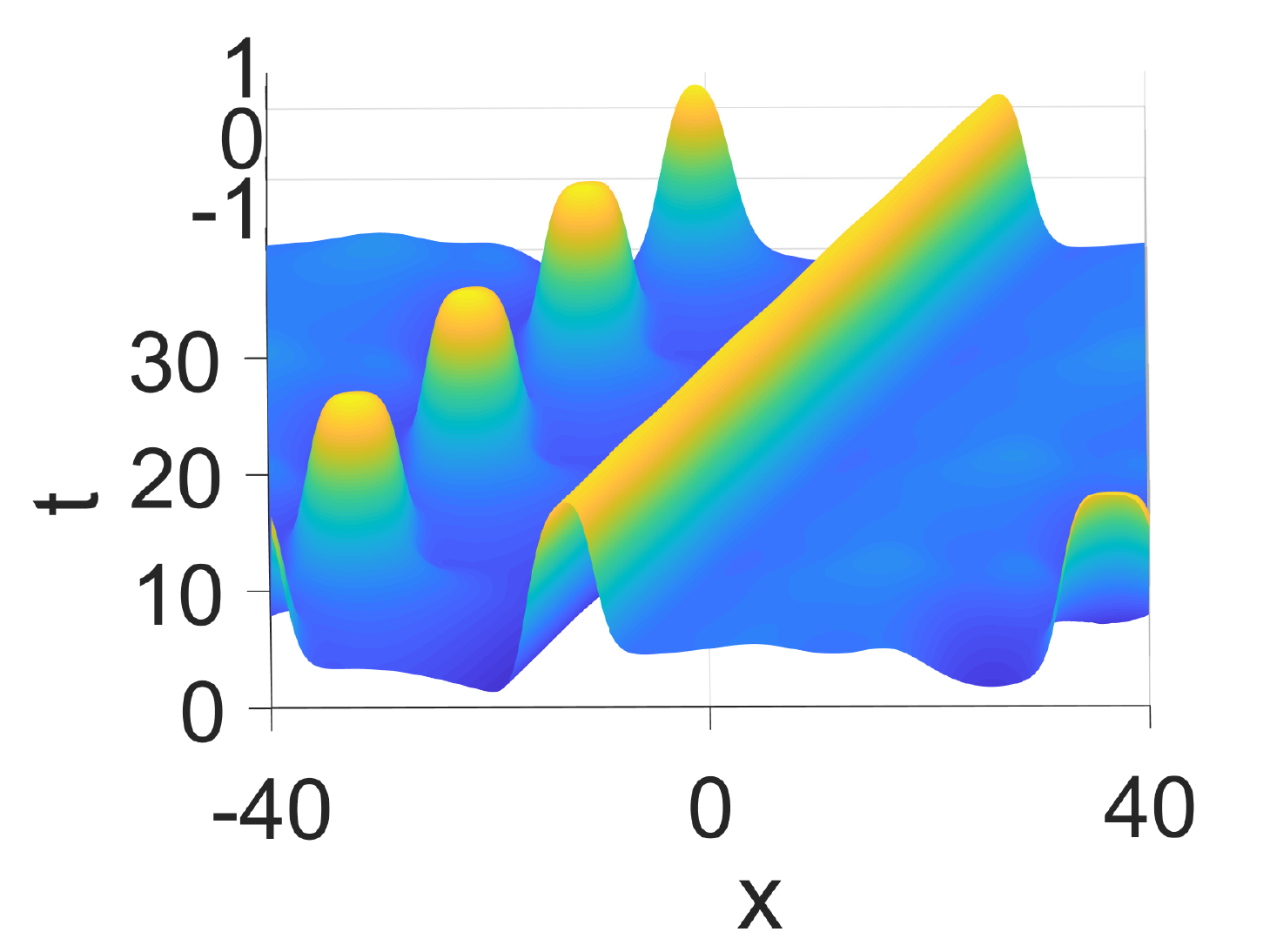}
		(b)\includegraphics[width=0.21\textwidth]{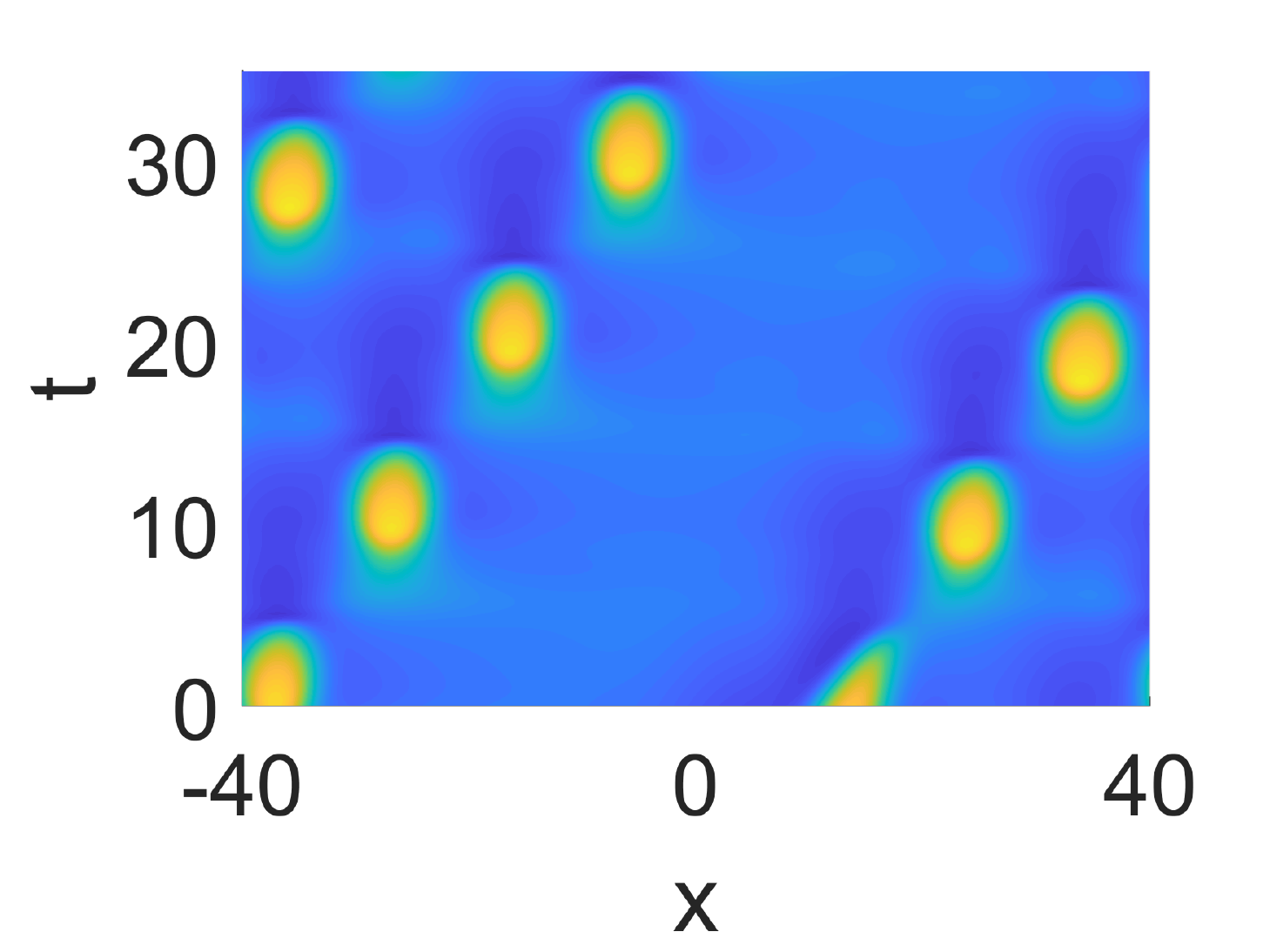}
		(e)\includegraphics[width=0.21\textwidth]{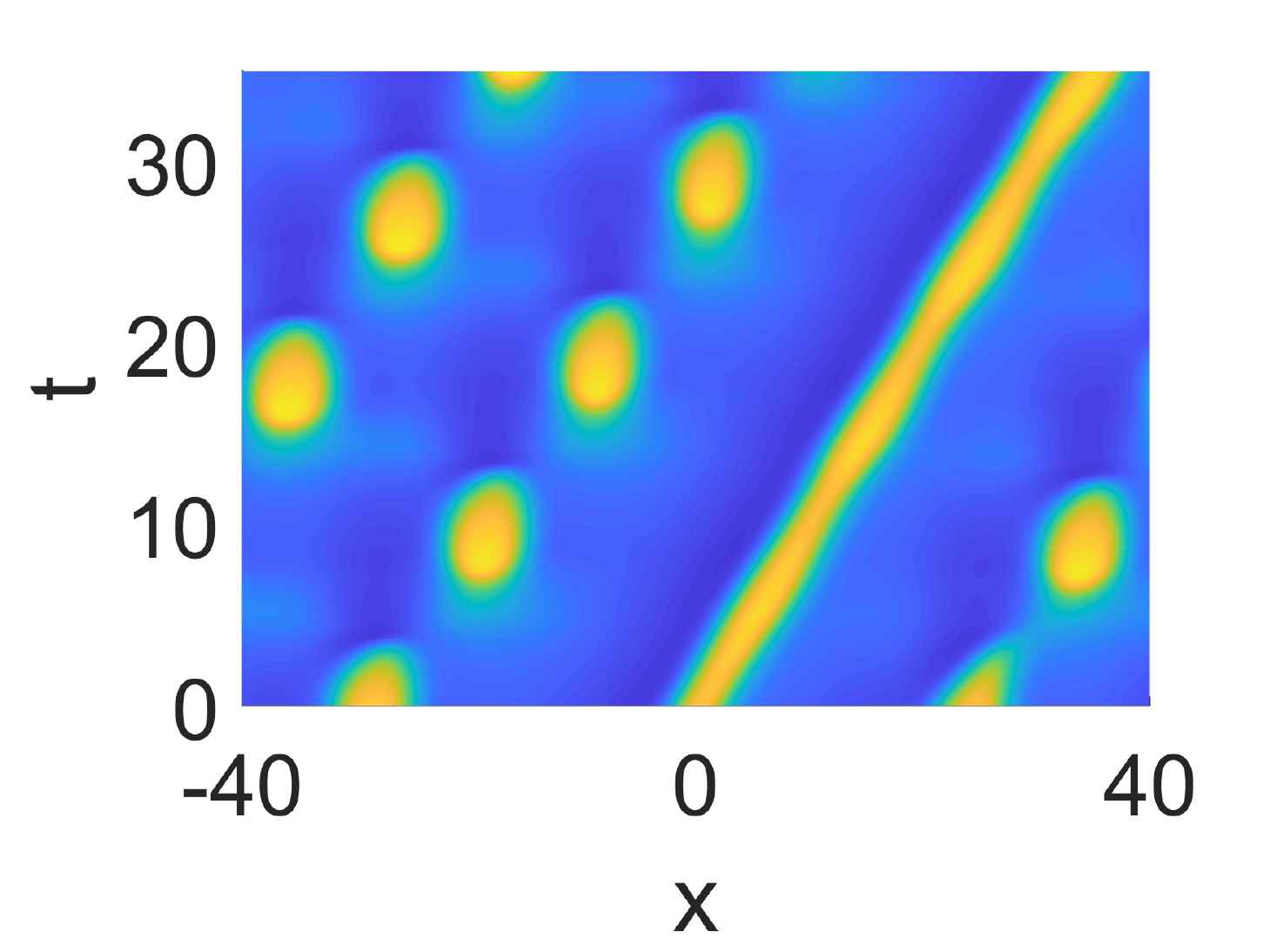}
		(c)\includegraphics[width=0.21\textwidth]{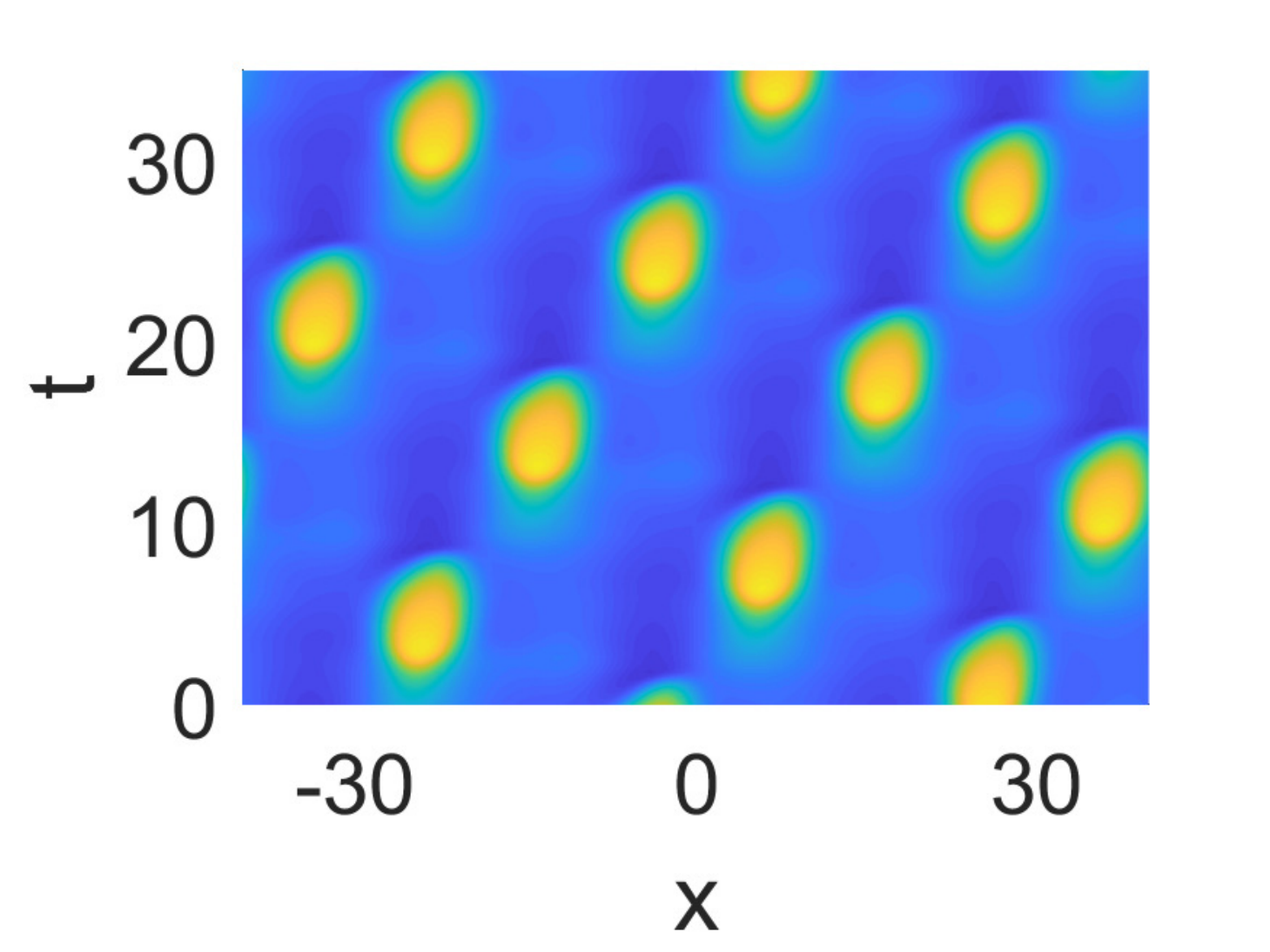}
		(f)\includegraphics[width=0.21\textwidth]{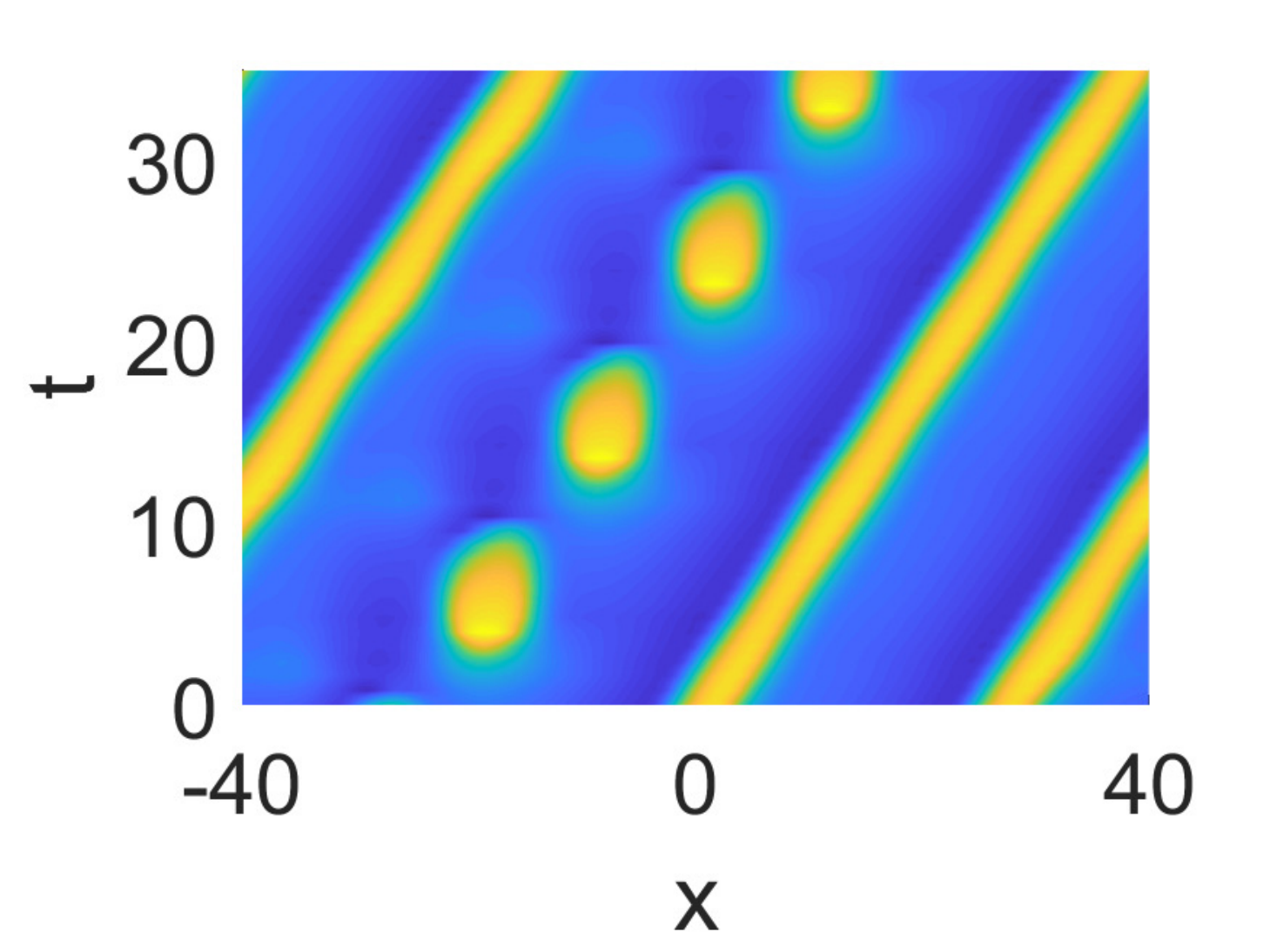}
		\caption{Space-time plots showing a selection of states obtained through DNS of \reff{bm0} with periodic boundary conditions. (a) JO at $k_1=-8.25$, (b) bound state of two JOs at $k_1=-9$, (c) 3JPP state at $k_1=-8.5$, (d) bound 1JO1TP state at $k_1=-7.8$, (e) 2JO1TP bound state at $k_1=-8.52$, and (f) 1JO2TP bound state at $k_1=-8.26$.  Other parameters are as in \reff{opar}. All plots show $u(x,t)$, with colorscale $-1.7$ to $1.5$.} 
		\label{nfa}
	\end{figure}
	
	In 2006,~\citet{YZE06} reported a {hitherto unseen spatially localized state they called a {\it jumping oscillon} (JO). Related behavior, a \textit{jumping wave}, was subsequently observed in the Belousov–Zhabotinsky reaction in a microemulsion~\cite{cherkashin2008discontinuously}.} JOs resemble the oscillons observed in parametrically driven systems~\cite{umbanhowar1996localized,lioubashevski1999oscillons} but translate at the same time, whereby they disappear and then reappear at a shifted location. The process repeats, generating a time-periodic state in a frame moving with the average translation speed, as shown in Fig.~\ref{nfa}(a). \citet{YZE06} and discovered these JOs in direct numerical simulations (DNS) of a three-variable FitzHugh--Nagumo (FHN) reaction--diffusion (RD) equation but provided little explanation of the origin of these states and the plethora of states associated with them. The simulations also show that these states can collide and either annihilate or combine in a uniformly translating localized pulse and that they can organize themselves into traveling rafts with a spatially periodic or crystalline structure. Similar structures have also been recently found in an active phase-field crystal model~\cite{ophaus2021} and in nonlinear optics~\cite{schelte2020dispersive}, implying broader applicability. Tailored patterns with spatiotemporal modulation, some of which are exhibited in Fig.~\ref{nfa}, can be envisioned as building blocks of information and storage handling~\cite{coullet2000stable} especially in the context of chemo-liquid
	computers~\cite{borresen2009neuronal,hiratsuka2009toward,adamatzky2011topics,gorecki2015chemical}.
	
	In this Letter, we focus on the origin and the bifurcation structure of stable and unstable JO branches, as well as traveling wave (TW) and traveling pulse (TP) branches. In view of the abundance of solutions of each type (Fig.~\ref{nfa}), numerical continuation must overcome possible branch jumping (undetected switching to a nearby branch in the predictor--corrector continuation setup). Nevertheless, the method enables us to identify the origin of the JOs through a modulational (Hopf) instability {of already unstable} TPs, and to obtain the bifurcation structure of accompanying multi-JOs [Fig.~\ref{nfa}(b)] and jumping periodic patterns (JPP) [Fig.~\ref{nfa}(c)] as well as the multitude of mixed JO and TP states [Figs.~\ref{nfa}(d-f)].\\ 
	
	\noindent \textit{Model equations and wave instability} -- We employ the Purwins system originally developed~\cite{schenk1997interacting} as a phenomenological model of an electrical discharge system exhibiting multiple stationary and moving localized states~\cite{purwins2010dissipative}. The system is a three-variable FHN system with one activator and two inhibitors acting on distinct timescales, and has broad applicability in studies of dissipative solitons in excitable RD systems in both one (1D) and two (2D) spatial dimensions (pulse interactions in 1D~\cite{doelman2009pulse,van2011planar,van2014bifurcations,teramoto2021traveling,nishiura2021matched} and time-dependent spots in
	2D~\cite{schenk1997interacting,or1998spot,bode2002interaction,gurevich2006breathing}). The closely related models studied
	in~\cite{VE04,nishiura2005scattering,yochelis2008generation,stich2009self,marasco2014vegetation,yochelis2015origin} exhibit similarly rich dynamics.
	
	The Purwins system in 1D reads 
	\begin{eqnarray}\label{bm0}
		\nonumber  \pa_t u&=&D_u\pa_x^2 u+k_1+k_2u-u^3-k_3v-k_4w,\\
		\tau \pa_t v&=&D_v\pa_x^2 v+u-v,\\
		\nonumber  \vt\pa_t w&=&D_w\pa_x^2 w+u-w, 
	\end{eqnarray} 
	where $k_4$, $\tau$, $\vt$ are parameters and $D_u$, $D_v$, $D_w$ are diffusion coefficients. The JOs in \cite{YZE06} were obtained with $(k_2,k_3,k_4,\tau,\vt,D_u,D_v,D_w)=(2,10,2,50,1,1,50,60)$ and near
	$k_1=-8.5$. Here we consider a similar parameter regime but with 
	\begin{equation}\label{opar}
		(\vt,D_u,D_v,D_w)=(0.5,2,25,100) 
	\end{equation}
	in order to separate out the branches in the bifurcation diagrams, and also employ $k_1$ as a control parameter. We supplement~\reff{bm0} with periodic boundary conditions on the 1D domain $x\in [-L/2,L/2]$, where $L=80$, and let $U\equiv(u,v,w)$. In our parameter regime, \reff{bm0} has a unique spatially homogeneous steady state $U_*=(u_*,u_*,u_*)$ (see SM~\ref{SM:disp}), which loses stability to a wave (finite wave number Hopf) instability at $k_1=k_{1c}\approx -7.6$ with a critical wave number $q=q_c$ associated with the critical wavelength $\lambda_c=2\pi/q_c \approx 20$; the dispersion relation at the onset
	is shown in SM~\ref{SM:disp}.
	
	\begin{figure*}[tp]
		\centering
		({\color{orange}{\large$\blacktriangle$}})\includegraphics[height=2.6cm]{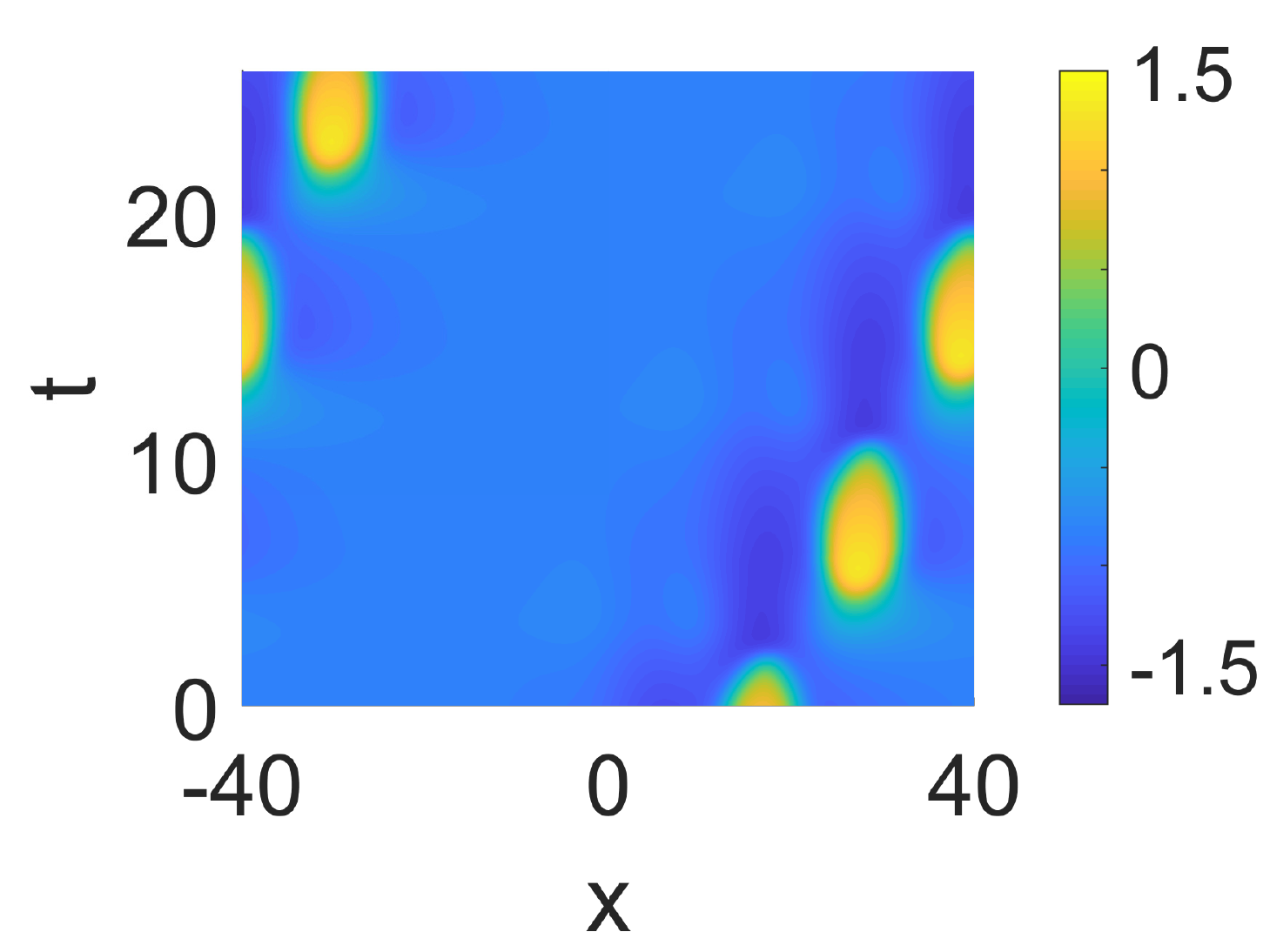}
		({\color{orange}{\small$\blacklozenge$}})\includegraphics[height=2.6cm]{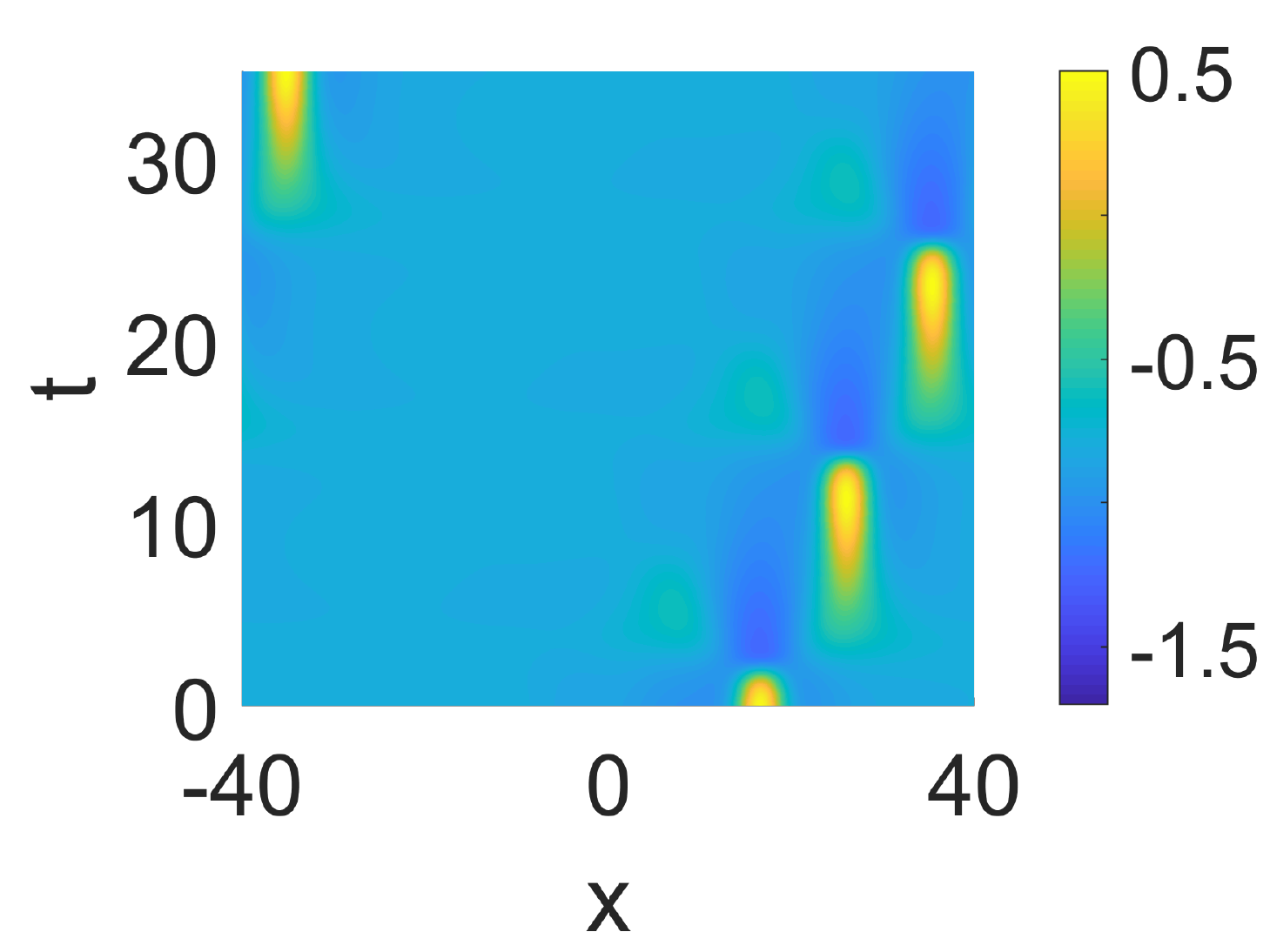}
		({\color{orange}{\small$\blacksquare$}})\includegraphics[height=2.6cm]{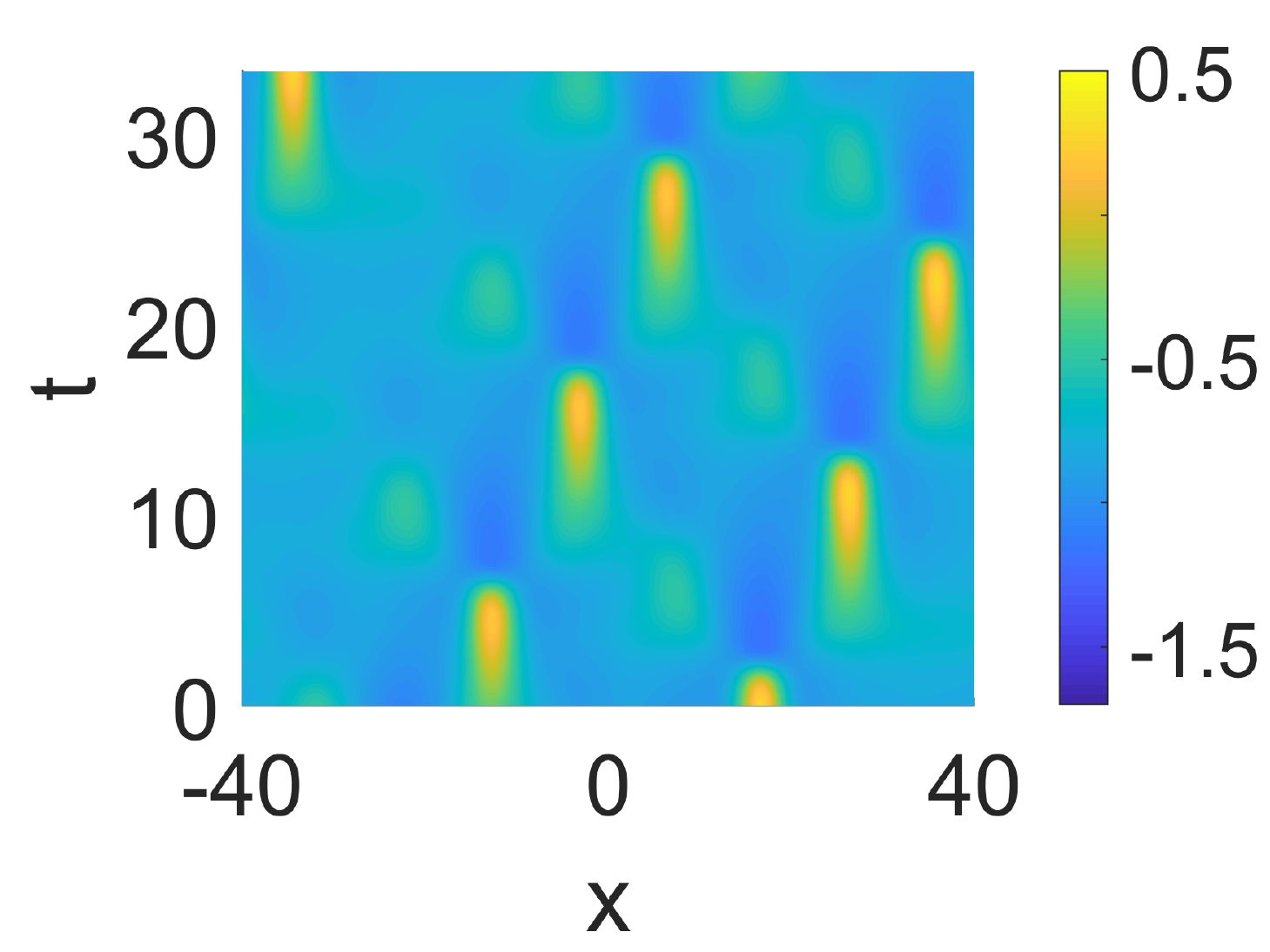}
		({\color{orange}{\large$\blacktriangledown$}})\includegraphics[height=2.6cm]{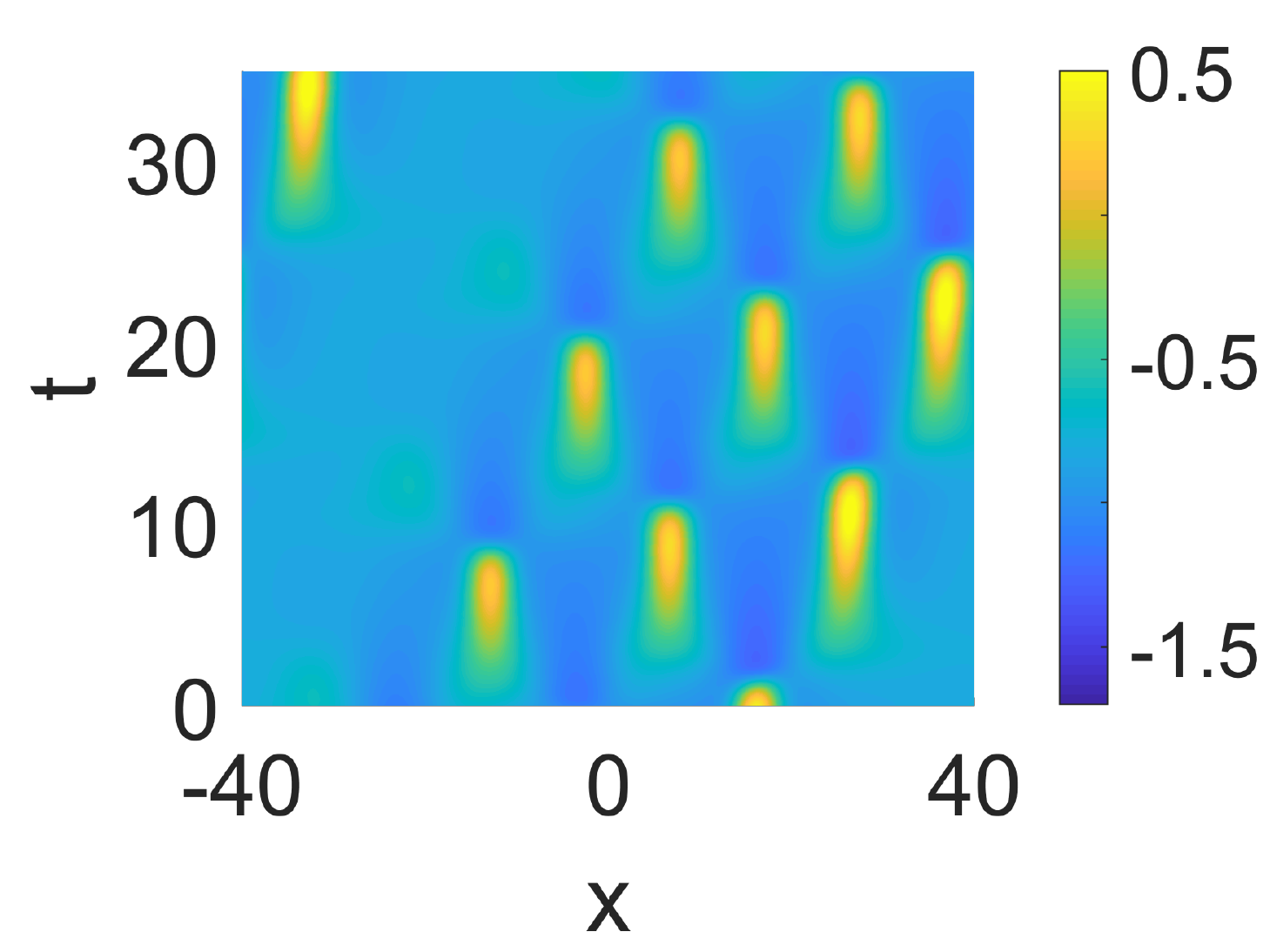}
		\includegraphics[width=0.9\textwidth]{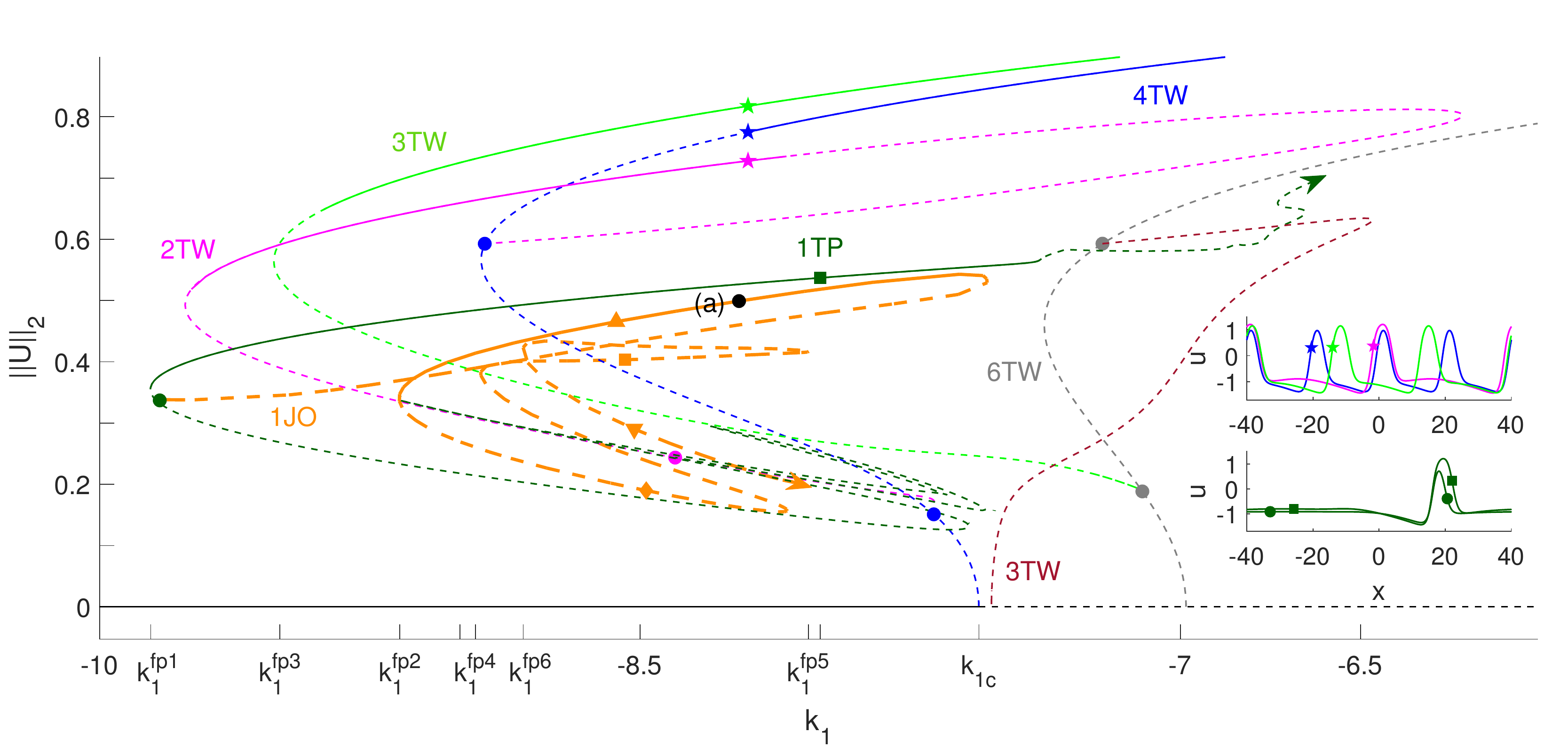}
		\caption{Partial bifurcation diagram showing $||U||_2$ as a function of $k_1$ for the uniform branch (black line), traveling waves (blue), traveling 1--pulse (green), time-modulated (including JO, orange) pulses, and further TW branches as indicated, with stability indicated by solid lines. The branch of modulated traveling pulses becoming JOs bifurcates from a Hopf point near $k_1{=}k_1^{\text{fp}1}$ (green bullet) and the JOs are stable for $k_1^{\text{fp}2}{<}k_1{\lesssim}k_{1c}$; the locations $k_1^{\text{fp}j}$ indicate other folds. Selected traveling (right) and time-modulated (top) solutions at different locations are also shown. The black bullet labeled (a) on the orange branch corresponds to the JO shown in Fig.~\ref{nfa}(a). Other parameters are as in \reff{opar}.} 
		\label{jof3}
	\end{figure*}
	
	The primary wave instability generates branches of spatially 
	periodic traveling and standing waves~\cite{knobloch1986oscillatory}, hereafter 4TW and 4SW, with four wavelengths in the domain (see SM~\ref{SM:disp}). Subsequent primary bifurcations generate 3TW, 5TW, 6TW and eventually 2TW. In what follows, we focus on traveling solutions that coexist with stable $U_*$, i.e.~on the \textit{excitable} regime, and show that the JOs are related to a temporal modulation of traveling solutions.\\ \\
	\noindent \textit{Continuation methodology} -- To compute solutions that are steady in a comoving frame (TW or TP), we rewrite (\ref{bm0}) in a reference frame moving with speed $s$ to the right, i.e.,~we add $-s\pa_x(u,v,w)$ to the right hand side of (\ref{bm0}), set the time derivatives to zero and solve the resulting nonlinear eigenvalue problem for $(U,s)$ using the phase condition
	\begin{equation}\label{se1} 
		\spr{\pa_x\Uold,U}\stackrel!=0, 
	\end{equation} 
	where $\Uold$ is the solution from the last continuation step, and 
	$\spr{f,g}\equiv\int_{-L/2}^{L/2}f(x)g(x)\dd x$. {This minimizes the 
		$L^2$ distance of the current step to translates of the previous step}~\footnote{\label{cfoot}{ {Additionally, $k_1$ is solved for in an arclength continuation setting, where the independent parameter is the arclength along a branch; this setting allows one to follow branches that oscillate back and forth in $k_1$, and in particular to pass folds.}}}. The continuation is thus orthogonal to the group orbit of translations $u(x{-}\xi)$, with $\pa_x$ as generator of the associated Lie algebra. We represent the traveling solutions using the norm $||U||_2\equiv\sqrt{L^{-1}\int [u(x)-u_*]^2\dd x}$, see the TW branches in Fig.~\ref{jof3}.
	
	For solutions of JO type or more generally modulated TW (mTW), we retain the time derivatives and solve for both the (mean) frame speed $s$ and the oscillation period $T$. To do so, we extend the phase condition to 
	\begin{equation}\label{avs} 
		q_H(U):=\sum_{i=1}^{m-1} \spr{\pa_x U_*, U(t_i)}\stackrel!=0, 
	\end{equation} 
	where $U_*=U_*(x)$ is a reference profile (usually $U_H(x)$, the spatial profile at the Hopf point) and
	$t_1, t_2\ldots, T$ are the grid points of the time discretization. Consequently, for mTW we have the three unknowns $(U,s,T)$ and solve the two equations \reff{bm0} and \reff{avs} together with the additional temporal phase condition
	\begin{equation}
		\int_0^T \spr{\ddt \Uold(t'),U(t')}\dd t\stackrel!=0, 
	\end{equation} 
	{to make the continuation orthogonal to the group orbit of time 
		translates.} 
	We also modify the norm $\|\cdot\|_2$ to 
	\begin{equation}\label{phase}
		\|U\|_2{=}\sqrt{\frac 1{TL} \int_0^T \int_{-L/2}^{L/2}
			\Bra{u(x,t){-}u_*}^2\dd x\dd t}.  
	\end{equation} 
	For the TW (TP) branches we thus have $n_u+2$ unknowns (including $k_1$, see \cite{Note1}) and $n_u+1$ equations, where $n_u=3n_p$ and $n_p$ is the number of spatial discretization points, while for the mTW (and JO) branches we have $mn_u+3$ unknowns (again including $k_1$) and $mn_u+2$ equations, where $m$ is the number
	of temporal discretization points. For our domain, we typically use $n_p\approx 1000$ discretization points in space and $m=50$ in time (for mTW), yielding $\approx 150000$ degrees of freedom. The predictor/corrector continuation method uses a corrector based on Newton's method and carries the danger of branch--jumping when many solutions are close together. To mitigate this, we monitor the convergence speed of the Newton loops, cf.~\cite[\S3.6]{p2pbook}. We also monitor selected eigenvalues of the linearizations to check stability and detect possible branch points, which are then localized, for subsequent branch switching 
	\footnote{\label{flfoot}For (relative) time--periodic orbits (PO), the role of eigenvalues (for stability, bifurcation detection, and branch switching) is played by Floquet multipliers. {These multipliers (in particular for PDE discretizations) may differ by many orders of magnitude and their (stable) numerical 
			computation therefore requires advanced methods such as periodic Schur decomposition 
			and hence is numerically expensive \cite[\S3.5]{p2pbook}}. Moreover, for some 
		points on our PO branches the multiplier computations do not converge. 
		We therefore refrain from computing bifurcations from POs, and instead check their stability {\it a posteriori} via DNS. See \cite{BBC15} for a review of the numerical methods related to POs in the presence of symmetry.}.\\
	
	\noindent \textit{Modulational instability of excitable pulses} -- Our first aim is to understand the origin of the JO states. Figure \ref{jof3} illustrates our results. The primary 4TW (blue) branch bifurcates subcritically from the primary Hopf point. Additional TW states are also shown: primary branches of 3TW (brown) and 6TW (gray), and two secondary TW branches that arise from period doubling, namely the 3TW
	(light green, from 6TW) and 2TW (pink, from 4TW). The dark green branch corresponds to TP states obtained by continuation from a stable 1TP state found in DNS starting from an initial condition obtained by cutting out a part of the 4TW solution (dark green square). The 1TP states are stable until the left-most fold at
	$k_1=k_1^{\text{fp}1}\approx -9.9$ where they lose stability. Beyond the second fold, near $k_1= k_{1c}$, the branch begins to {\it snake} \cite{knobloch2015spatial} forming 2TP and then 3TP states. The latter fail to connect to a 3TW state and begin to snake downwards until they connect at the pink bullet to the 2TW branch. In the opposite direction the 1TP undergo complex behavior before terminating back on the same 2TW branch (SM~\ref{SM:TP}). Thus, both the {1TP} branch and the 2TW branch represent branches that start and end on the same branch in secondary bifurcations. 
	
	{The properties of the JOs are closely related to these background states. We find that the JO branch (orange) emerges from the first Hopf bifurcation located on the \textit{unstable} portion of the 1TP branch (green bullet in Fig.~\ref{jof3}). Consequently, the JOs start out as an unstable \textit{small amplitude modulation} of the 1TP state, and turn into stable, fully developed JOs (orange triangle in Fig.~\ref{jof3}) only after the fold near $k_1=k_{1c}$, before losing stability again at the next fold at $k_1=k_1^{\text{fp}2}$}; the time $t=0$ state at the black bullet (a) was used as initial condition for the DNS in Fig.~\ref{nfa}(a). The stable JOs have the highest travel speed $s$ and shortest oscillation period $T$ along the whole orange branch (SM~\ref{SM:speed}).
	
	As the branch snakes below the fold at $k_1=k_1^{\text{fp}2}$, the solutions become unstable (orange diamond in Fig.~\ref{jof3}), before turning into a (smaller amplitude) 2JO bound state (at $k_1\approx k_1^{\text{fp}4}$,
	orange square in Fig.~\ref{jof3}), and then into a (yet smaller amplitude) 3JO bound state (orange down-triangle in Fig.~\ref{jof3}). Beyond this point, the continuation becomes unreliable in the sense that different numerical settings (finer discretizations/tolerances) may lead to different behavior, but the
	branch continues. \\
	
\begin{figure}[tp]
		\centering
		({\small$\blacklozenge$})\includegraphics[height=2.6cm]{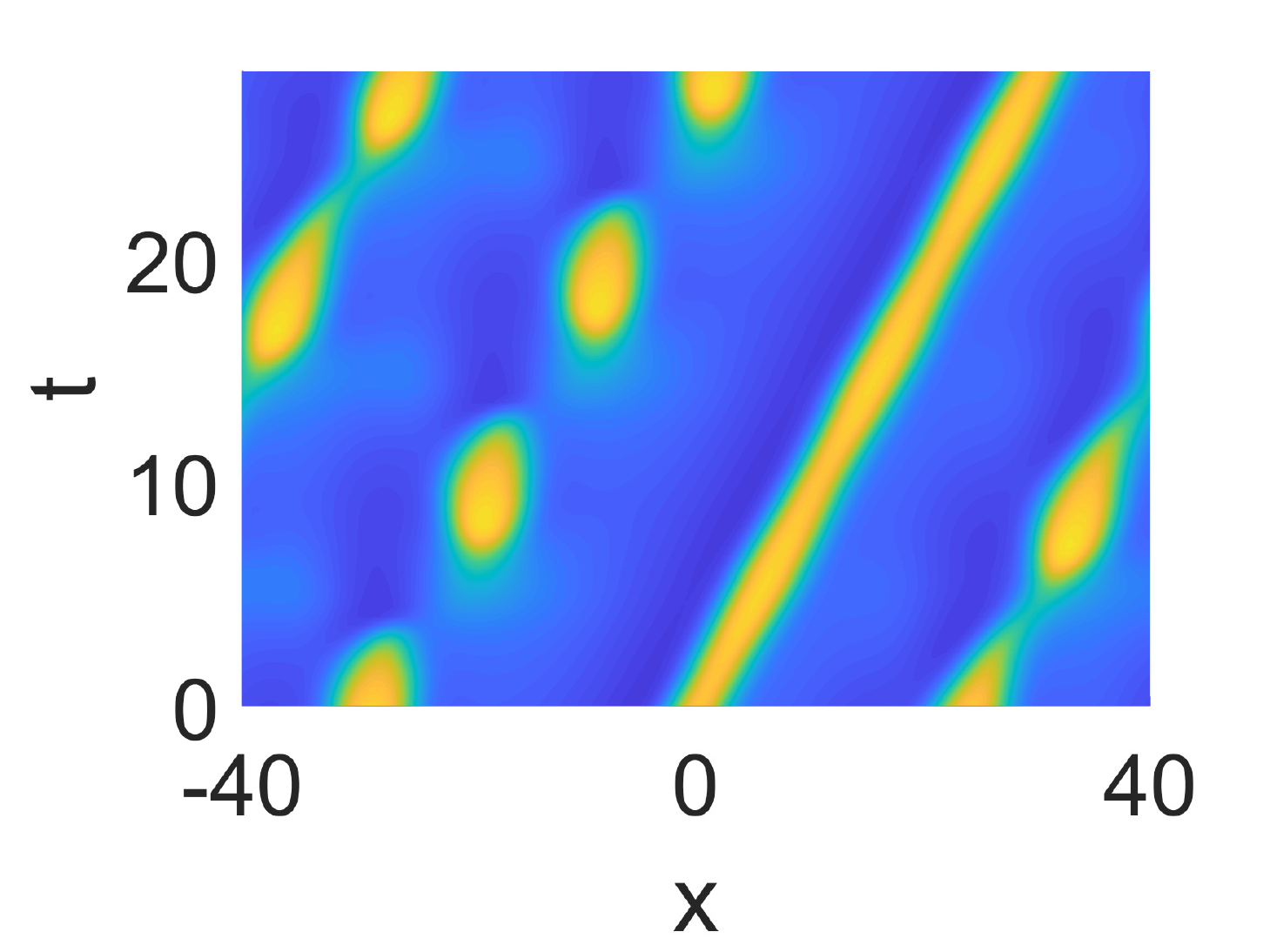}
		({\small$\blacksquare$})\includegraphics[height=2.6cm]{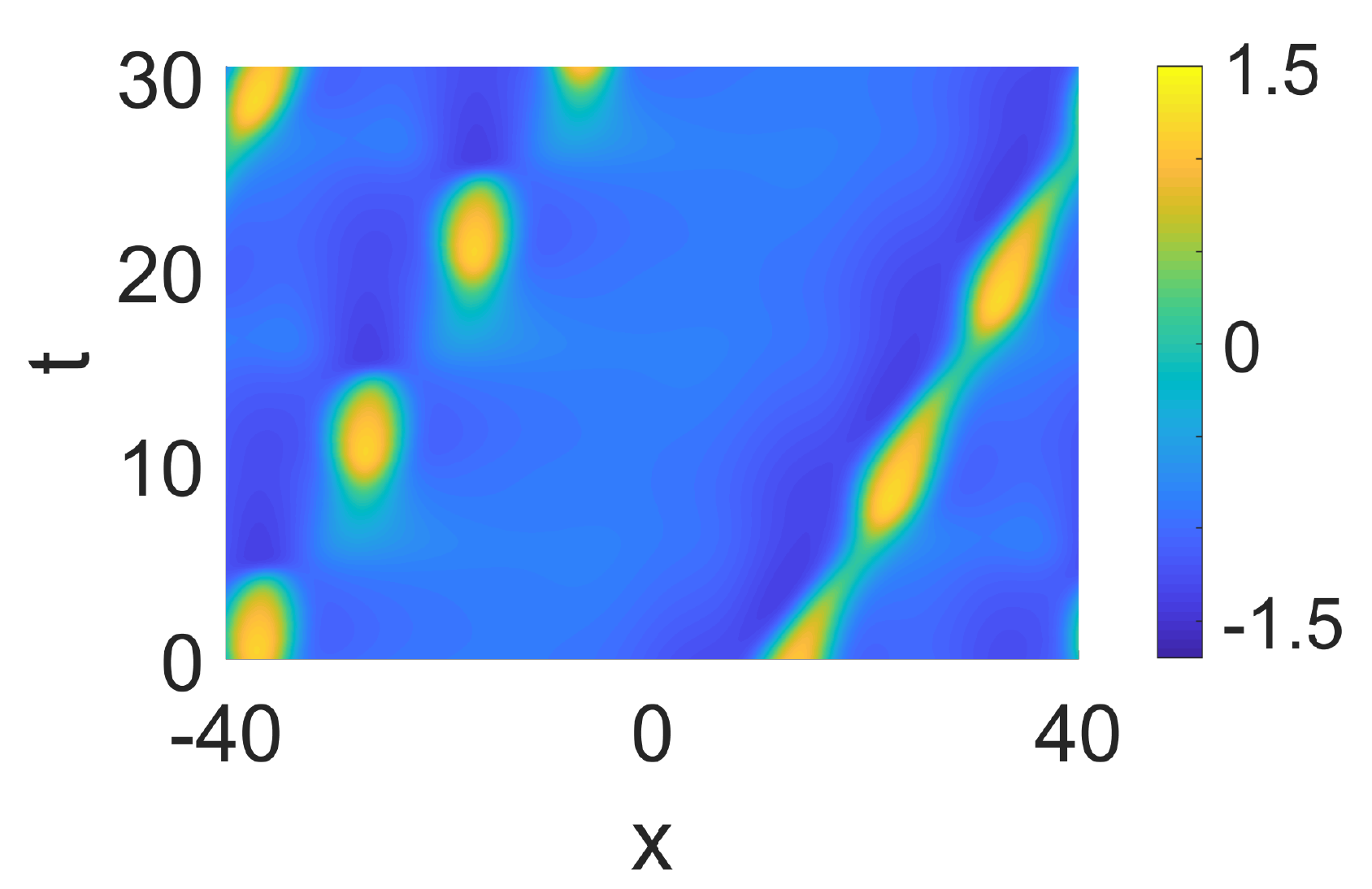}
		\includegraphics[width=0.95\columnwidth]{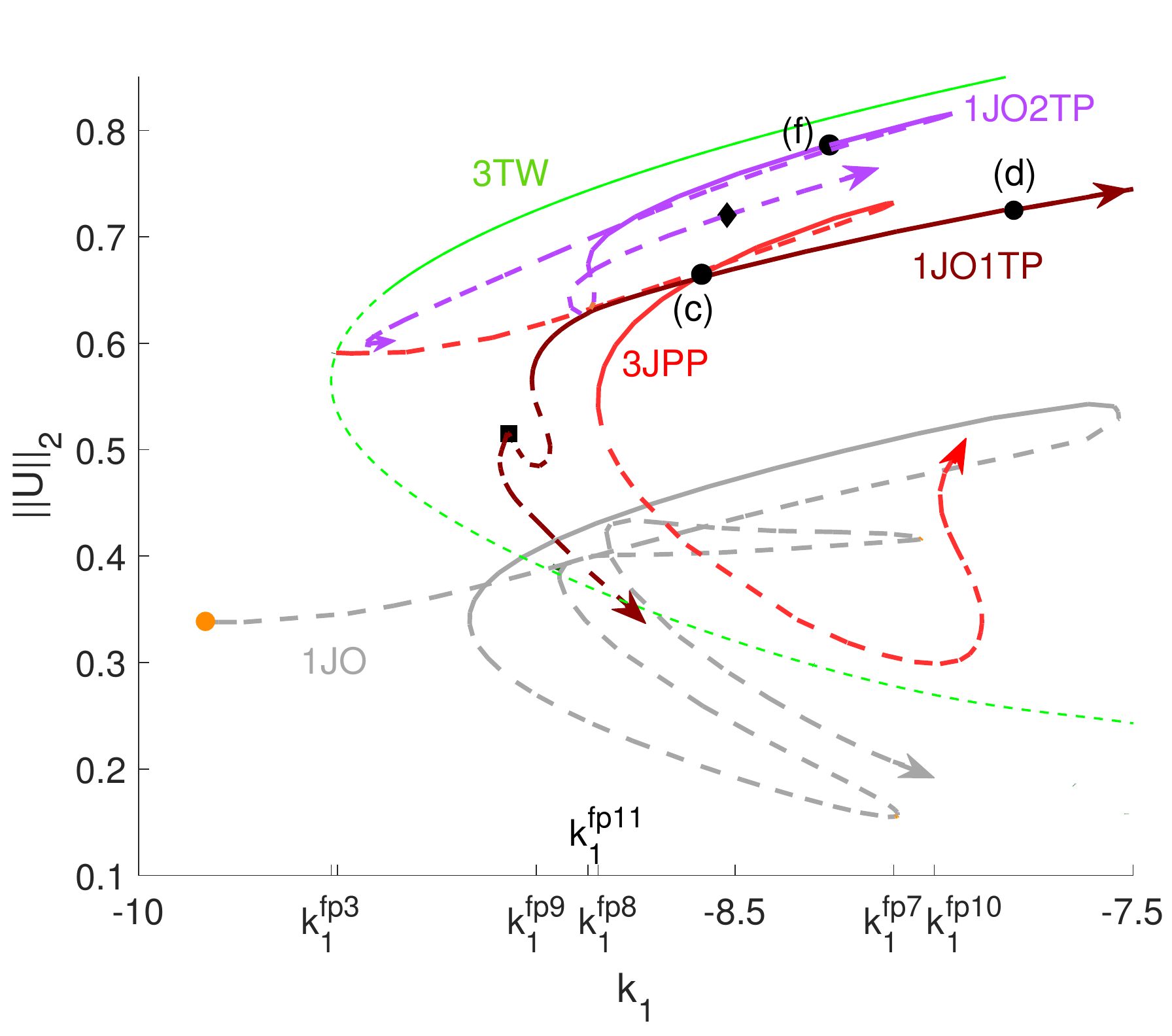}
		\caption{Partial bifurcation diagram showing $||U||_2$ for the 3TW branch (green), the associated 3JPP branch (red) and branches of bound states 1JO1TP (brown) and 1JO2TP (purple), with the JO branch kept in the background in gray; solid lines indicate stability. The 3JPP branch bifurcates from the Hopf point near $k_1=k_1^{\text{fp}3}$; other $k_1^{\text{fp}j}$ points mark folds designating the limits of stability. Selected time-modulated solutions at different locations are shown in the top panels (square/diamond symbols) and labeled bullets correspond to states shown in Fig.~\ref{nfa}. Other parameters are as in \reff{opar}.}\label{jof4}
	\end{figure}

	\noindent\textit{Bound states and mixed states} -- In Fig.~\ref{jof4}, we plot several branches that are associated with multi-JO solutions and mixed JO/TP states. The starting point for finding these states is a domain-filling 3JPP state (red branch in Fig.~\ref{jof4}). These extended 3JPP states are characterized by a pronounced phase gradient and bifurcate from a 3TW (light green) branch via a Hopf bifurcation just above $k_1=k_1^{\text{fp}3}\approx -9.5$ and are initially unstable. Stable states are found on the segment between the first two folds, $k_1^{\text{fp}8}<k_1<k_1^{\text{fp}7}$. Stability is again checked using DNS, yielding for instance Fig.~\ref{nfa}(c). There are many other Hopf points along the 1TP branch, along the 3TW branch, and along similar branches, which give rise to further mTP branches. Many of these are pairwise connected via small-amplitude mTW branches. We show one example of such a branch in SM~\ref{SM:mTW}.
	
	The overall picture of localized and extended JO states is quite robust with respect to parameter changes, although details of the connections between branches depend sensitively on parameters, as shown in SM~\ref{SM:mTW} for $D_v<25$. One quantitative effect of decreasing $D_v$ is the loss of the first two Hopf points on the 1TP branch (see SM~\ref{SM:mTW} for details), resulting in a shift of the Hopf point from which the JO branch originates.
	
	Figure~\ref{jof4} demonstrates that the continuation method can also be used to identify the bifurcation structure of more complex states, including mixed JO/TP states. Starting from (stable) bound states of JO and TP obtained from DNS, e.g., 1JO1TP [Fig.~\ref{nfa}(d)] and 1JO2TP [Fig.~\ref{nfa}(f)] and identifying the period $T$ and frame speed $s$ we can use continuation to compute both stable and unstable regimes associated with such solutions. Stable 1JO1TP (brown branch) states exist for $k_1>k_1^{\text{fp}8}$ while stable 1JO2TP (purple branch) are present between $k_1^{\text{fp}11}<k_1<k_1^{\text{fp}10}$. Moreover, unstable states (see the square and diamond symbols in Fig.~\ref{jof4}) provide initial conditions that converge, via DNS, to stable 2JOs [Fig.~\ref{nfa}(b)] and 2JO1TP [Fig.~\ref{nfa}(e)], respectively. The continuation of these states (not shown) follows the same procedure as used for single 
	JOs. \\
	
	\noindent\textit{Discussion} --
	We have shown that a successful understanding of the origin of the 1JO and associated states, such as the 2JO and 1JO1TP bound states or the JPP state discovered in~\cite{YZE06}, can be achieved within a careful continuation/bifurcation setting -- the only currently existing technique for such purposes. For the three-variable FHN model~\reff{bm0}, the continuation of excitable states turned out to hold significant challenges: while for the TW and TP we only have spatial degrees of freedom to deal with, for the mTP and mTW, as relative time-periodic orbits, we need a fine temporal resolution as well, leading to $10^5$ (and more) necessary degrees of freedom, even on the small domain used here. We have checked that our results persist to larger domains with more turns on the 1TP branch as it snakes (not shown), and similar behavior of the JO state, although some differences are inevitably present. The numerical difficulties mount, however, due to the abundance of states and possible branch jumping, issues that (mostly) do not arise on smaller domains or in DNS. We showed here that the origin of stable JOs is highly subtle, requiring first and foremost an understanding of the underlying 1TP states and their stability, but that our approach is up to the task. {Moreover, 2D DNS in the corresponding parameter range yield target-like jumping waves similar to those observed experimentally in the Belousov–Zhabotinsky reaction in a microemulsion~\cite{cherkashin2008discontinuously}, as shown in SM~\ref{SM:JW}. Thus our approach yields} useful insight into the extremely rich solution structure of \reff{bm0} in the excitable regime, and in particular into the origin of JOs and their subsequent snaking forming 2JOs, 3JOs,..., and ultimately domain-filling JPP arrays.
	
	Like the wide interest stimulated by the discovery of oscillons~\cite{umbanhowar1996localized,lioubashevski1999oscillons,blair2000patterns}, we expect that the methodology developed here for JOs will also lead to new theoretical questions as well as potential applications in other multi-variable excitable RD media. The rich yet programmable JO patterns strengthen the suggestion in~\cite{coullet2000stable} that localized states could be useful for data storage in computers with RD kinetics~\cite{borresen2009neuronal,hiratsuka2009toward,adamatzky2011topics,gorecki2015chemical}.
	\\ \\
	\noindent {\bf Acknowledgement}. We thank Svetlana Gurevich (M\"{u}nster) for helpful discussions. This work was supported in part by the National Science Foundation under Grant No. DMS-1908891 (EK).

\clearpage
\renewcommand\thefigure{S\arabic{figure}}
\renewcommand\theequation{S\arabic{equation}}
\renewcommand\thesection{S\arabic{section}}
\setcounter{figure}{0}
\setcounter{equation}{0}
\onecolumngrid
\section*{{\Large SUPPLEMENTARY MATERIAL}}

\section{Dispersion relation and wave instability}\label{SM:disp}
\begin{figure}[h!]
	\centering
	(a)\includegraphics[width=0.45\textwidth]{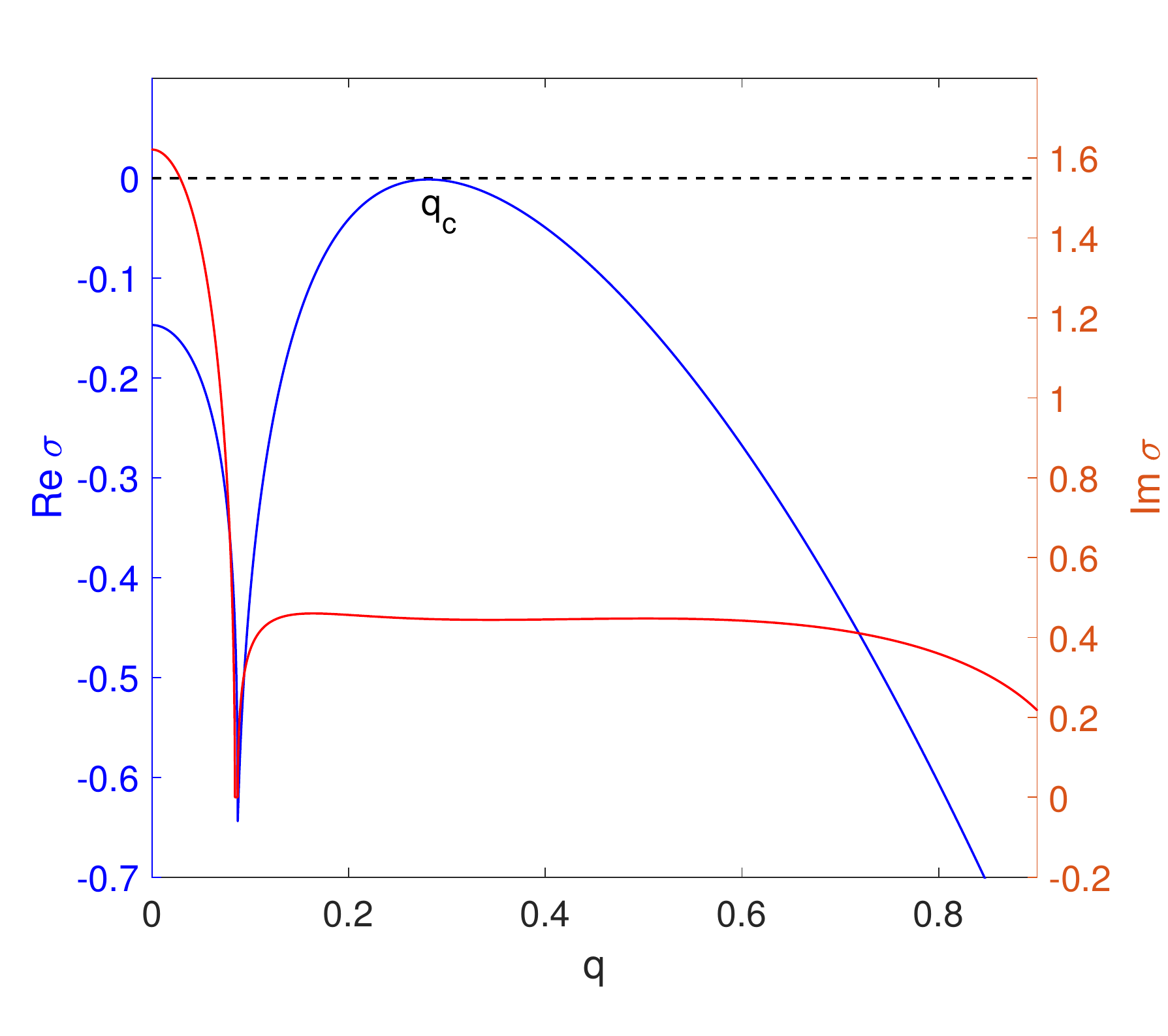}
	(b)\includegraphics[width=0.45\textwidth]{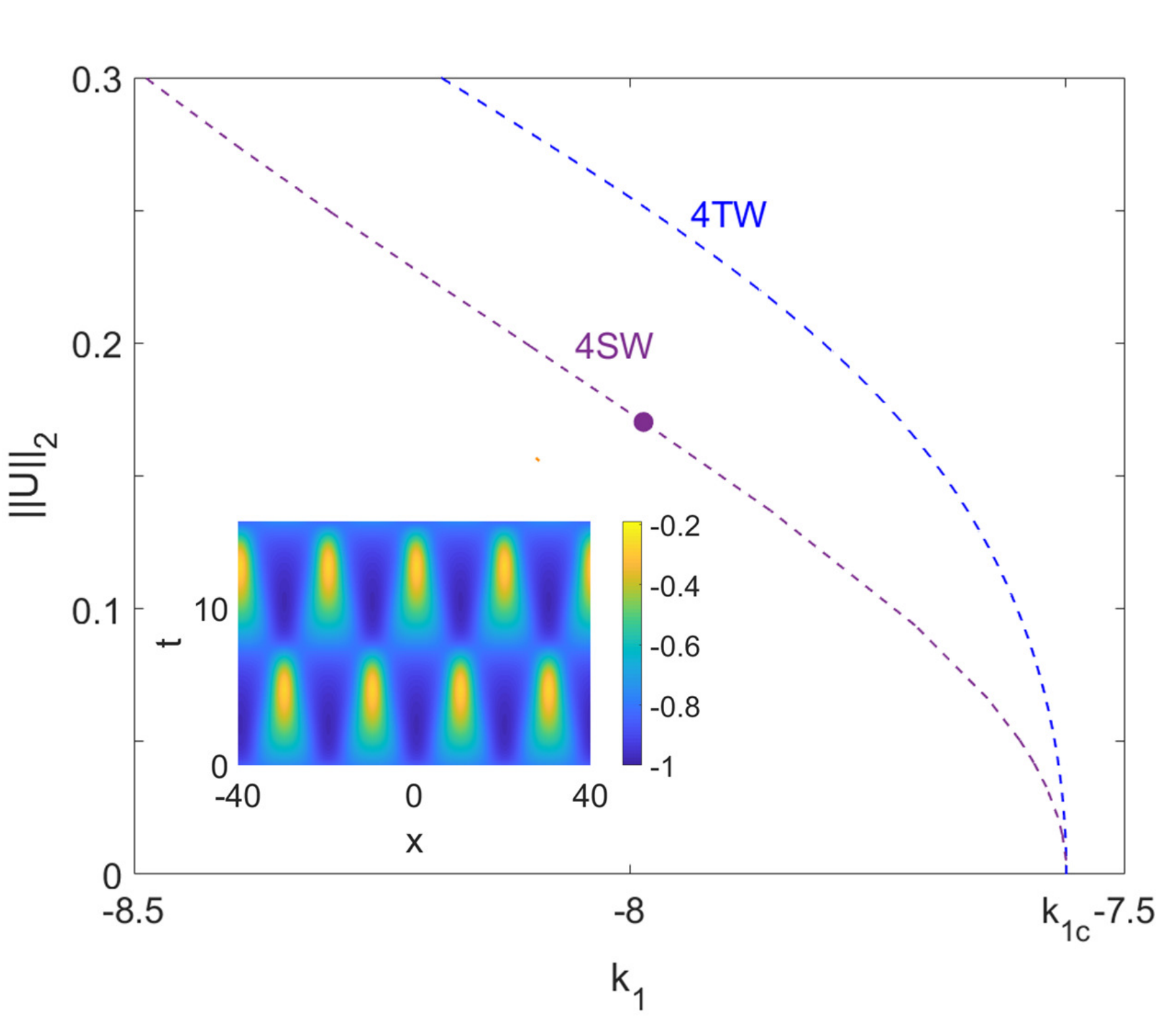}
	\caption{(a) Dispersion relation plotted at $k_1=k_{1c}\approx -7.59$. (b) Bifurcating branches of 4TW (blue) and 4SW (purple) from the Hopf onset with wavelength $\lambda=20 \approx 2\pi/q_c$ on a domain of size $L=80$. The inset shows a space-time plot of $u(x,t)$ corresponding to a 4SW state at $k_1=-8$, as indicated by the bullet.}
	\label{fig:disp}
\end{figure}

\noindent
The spatially homogeneous steady state is $U_*=(u_*,u_*,u_*)$~\cite{YZE06}, where
\[ 
u_*=\sqrt[3]{\del+k_1/2}-\sqrt[3]{\del-k_1/2},
\] 
with $\del=\sqrt{(k_1/2)^2+(p/3)^3}$ and $p=k_3+k_4-k_2$.
Linear stability of $U_*$ is computed numerically from the dispersion relation associated with the ansatz
	$   U-U_*\propto \exp [\sigma t - iqx],$ 
where $\sigma$ is the temporal growth rate corresponding to wave number $q$. In Fig.~\ref{fig:disp}(a), we show the dispersion relation at $k_1=k_{1c}$. The state $U_*$ is linearly stable if $\re\sig(q)<0$ for all $q\in\R$. 

In the parameter regime considered, $U_*$ is linearly stable for sufficiently negative $k_1$, and the first instability sets in at $k_1=k_{1c}\approx -7.6$ as $k_1$ increases and is of wave (finite wave number Hopf) type. At $k_{1c}$ we have $\re\sig(q\neq q_c)<0$, with $\re\sig(q_c)=0$ and $q=q_c\approx 0.285$. Thus, $q_c$ is the critical wave number at the onset with corresponding critical frequency $\im\sig(q_c) \approx 0.45$. The resulting bifurcation is a Hopf bifurcation with $O(2)$ symmetry, which gives rise to families of standing and traveling solutions (SW and TW, respectively), both of which bifurcate subcritically (i.e., in the direction of stable $U_*$), and are therefore initially unstable \cite{knobloch1986oscillatory}, as shown in Fig.~\ref{fig:disp}(b).

\section{Complete branch of traveling excitable pulses}\label{SM:TP}
\begin{figure*}[h!]
	\centering
	\includegraphics[width=0.9\textwidth]{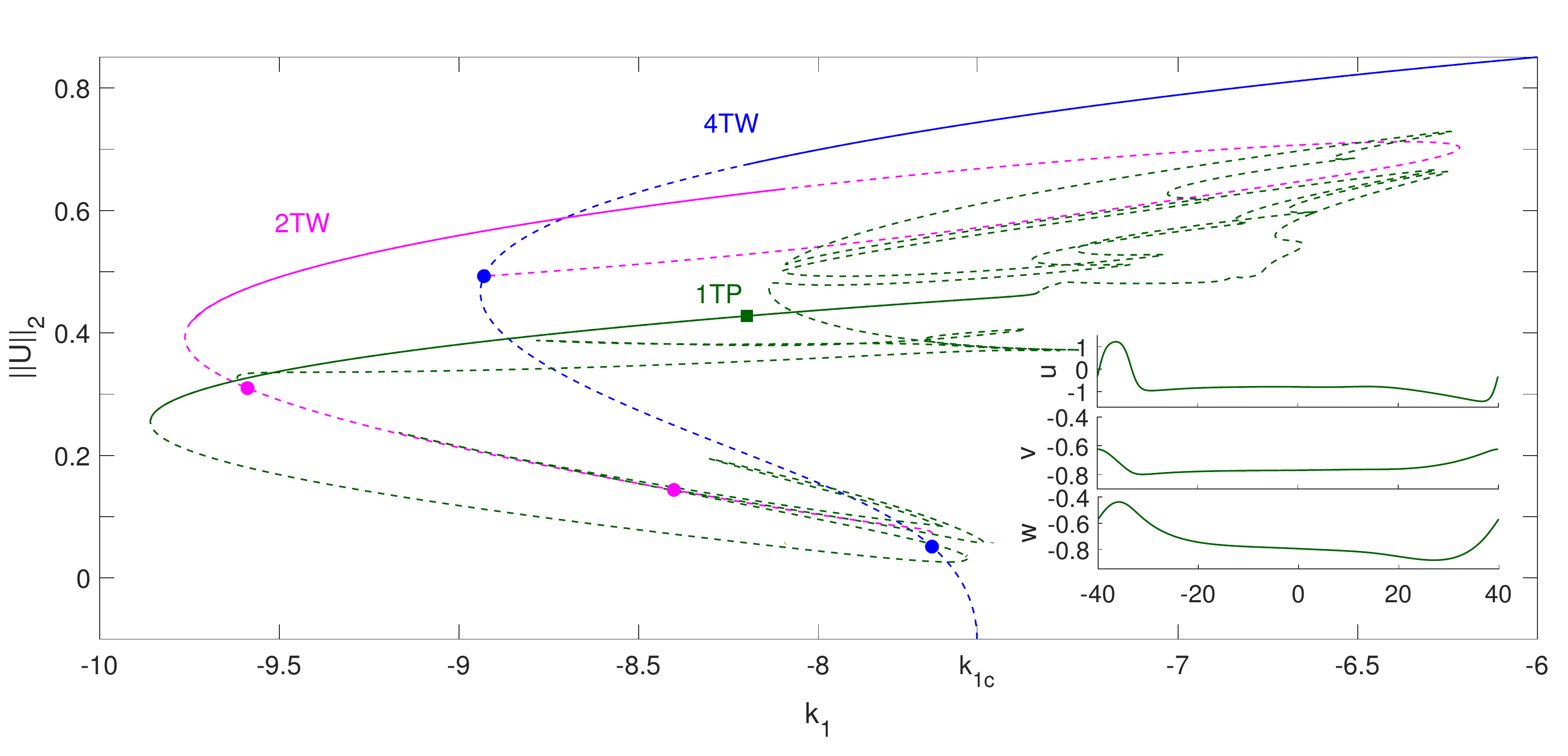}
	\caption{Complete 1TP branch, together with the 2TW and 4TW branches. 
		Solid lines mark stable regions and bullets indicate the branching points 
		(see main text for details). The profile of a traveling excitable pulse at
		$k_{1}=-8.2$ (green square) is shown on the right in all three variables.
		The color scheme is as in Fig.~2 of the main text.}
	\label{fig:TP}
\end{figure*}

In the main text (Fig.~2), we showed a partial branch of traveling pulses 
(1TP). The complete branch is shown in Fig.~\ref{fig:TP}. 

\section{Translation speed and temporal oscillation period of jumping oscillons}\label{SM:speed}	
\begin{figure}[bp]
	\centering
	(a)\includegraphics[width=0.45\textwidth]{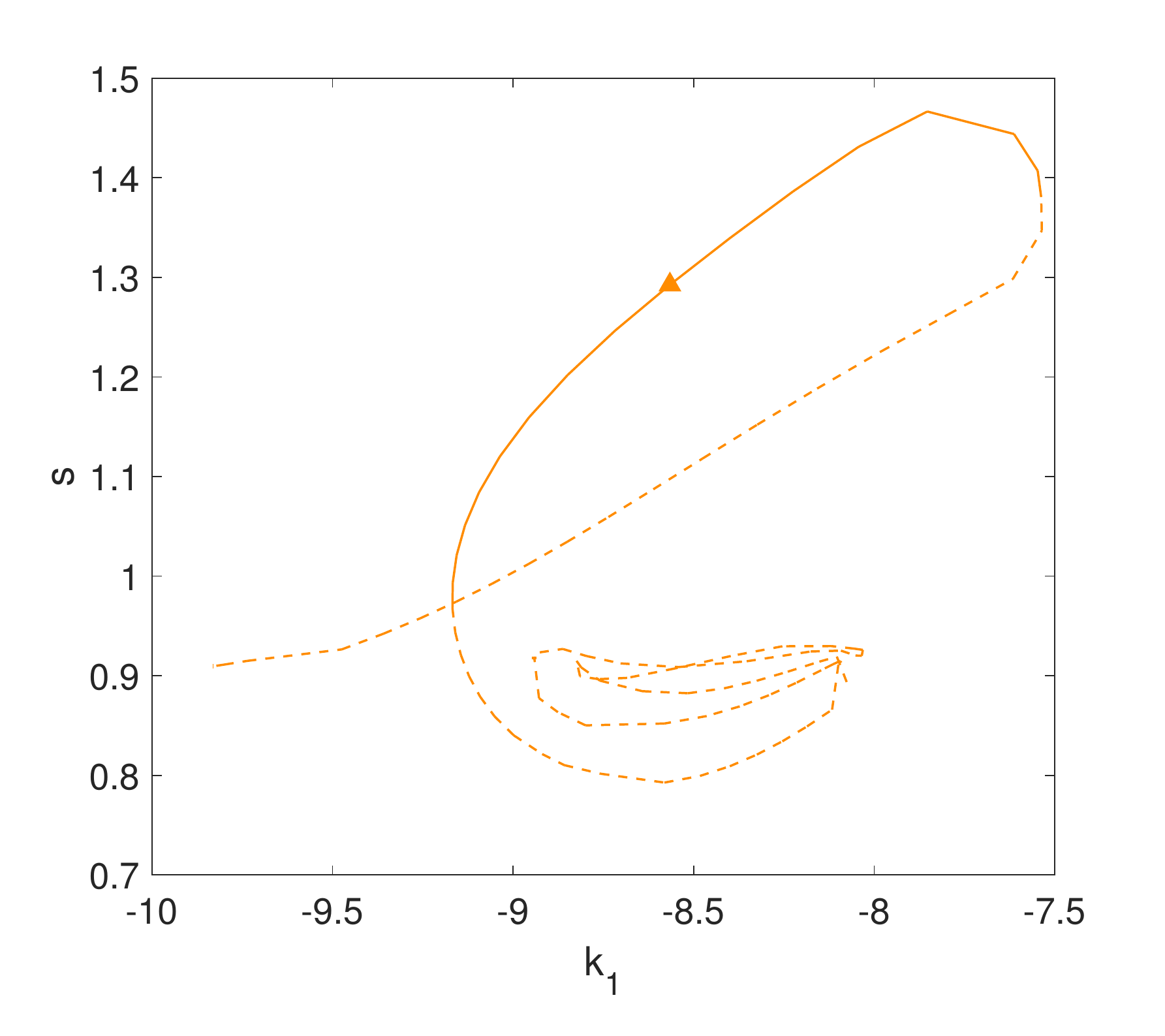}
	(b)\includegraphics[width=0.45\textwidth]{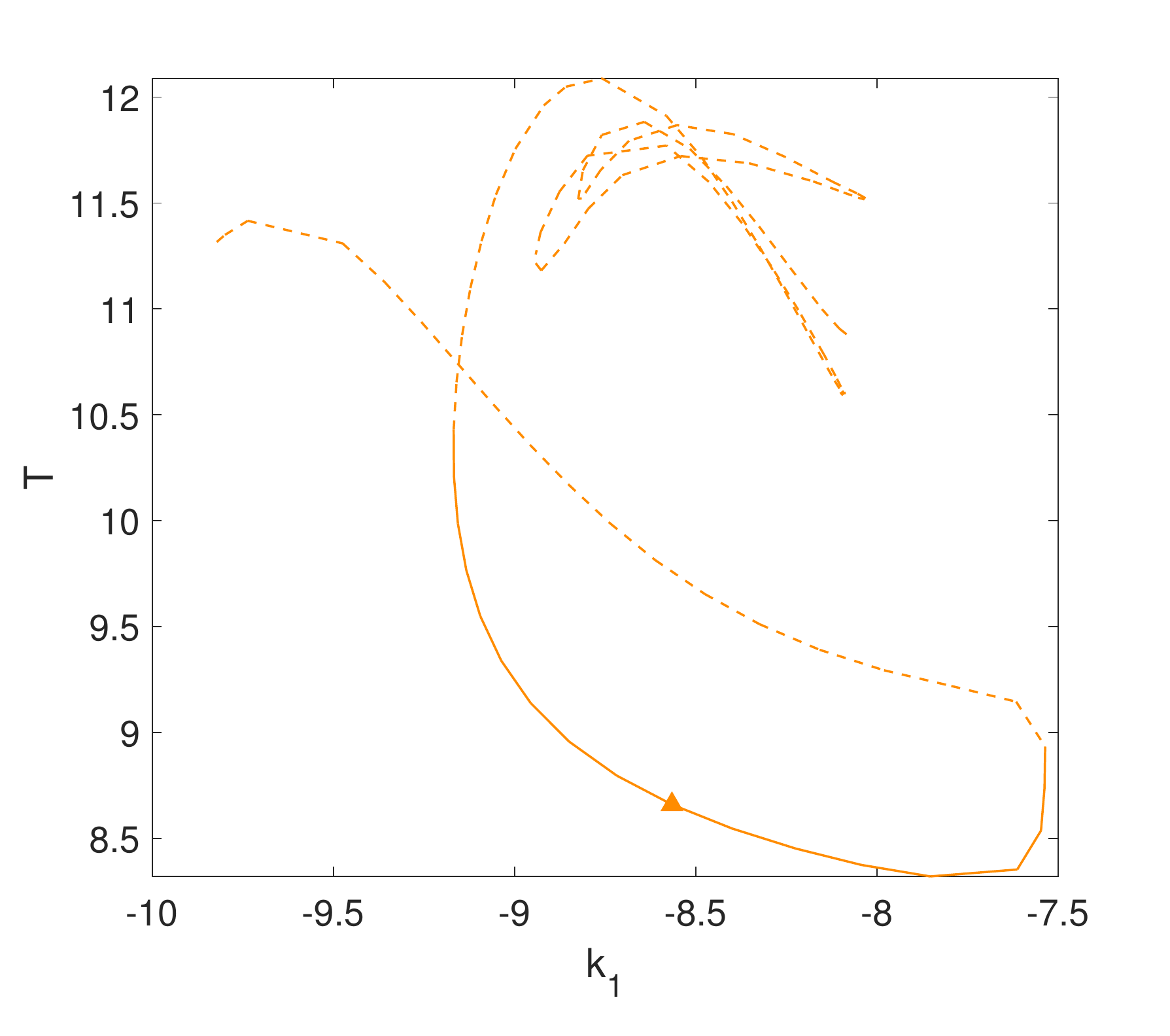}
	\caption{Branch of 1JO solutions shown in terms of (a) the mean translation speed $s$ and (b) the oscillation period $T$ along the branch, both as a function of $k_1$. Solid lines mark stable regions and the triangle marks the location of stable JO shown in Fig.~1(a) of the main text.}
	\label{fig:1JO}
\end{figure}

In the main text (Fig.~2), we showed the 1JO branch in terms of the $||U||_2$ norm. In Fig.~\ref{fig:1JO}, we complement this by showing the same 1JO bifurcation diagram in terms of (a) the mean translation speed $s$ and (b) the oscillation period $T$ in the moving frame.

\section{Other Hopf bifurcations of excitable pulses to modulated traveling waves}\label{SM:mTW}	

\begin{figure}[h!]
	\centering
	(a)\includegraphics[width=0.45\textwidth]{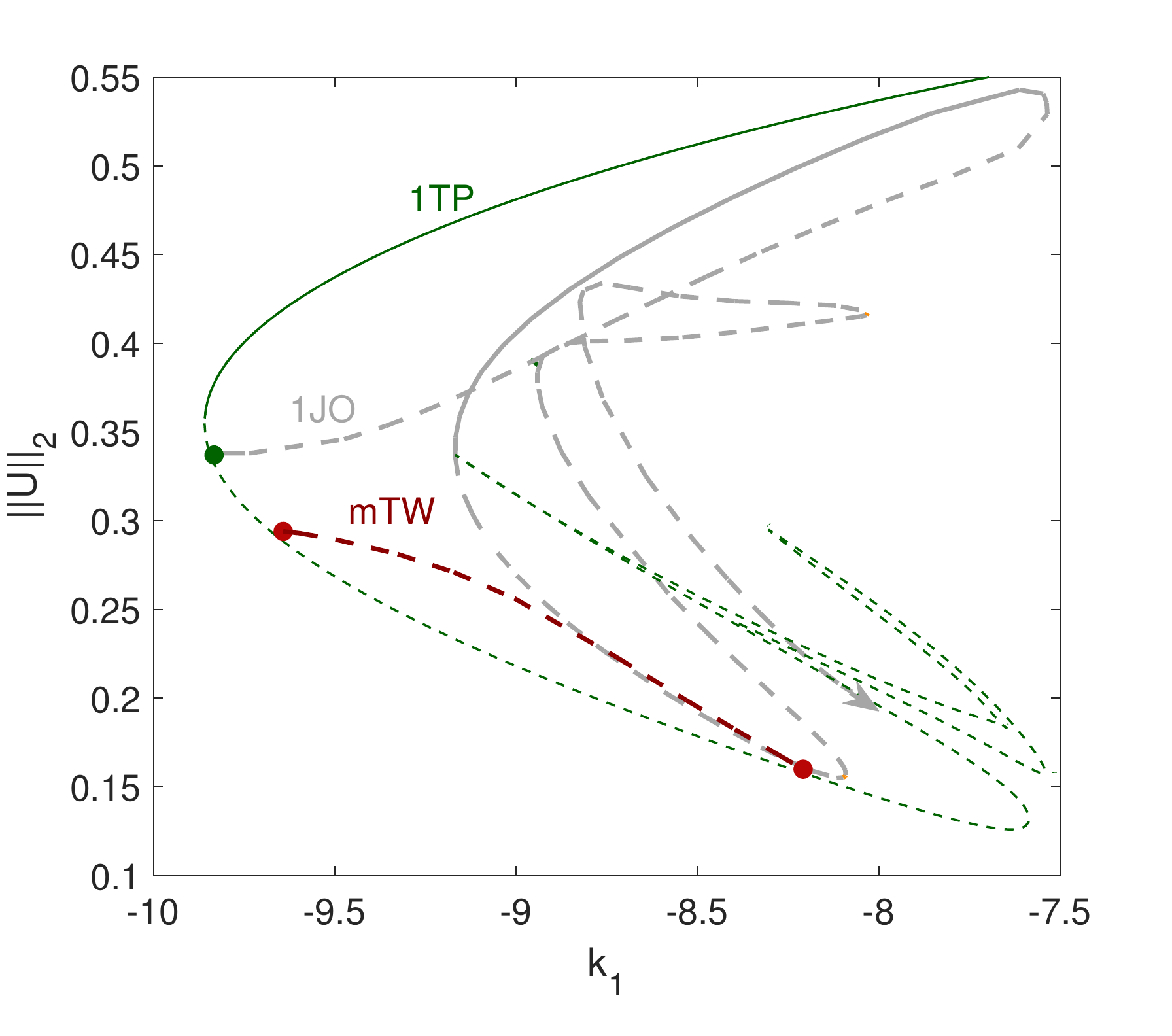}
	(b)\includegraphics[width=0.45\textwidth]{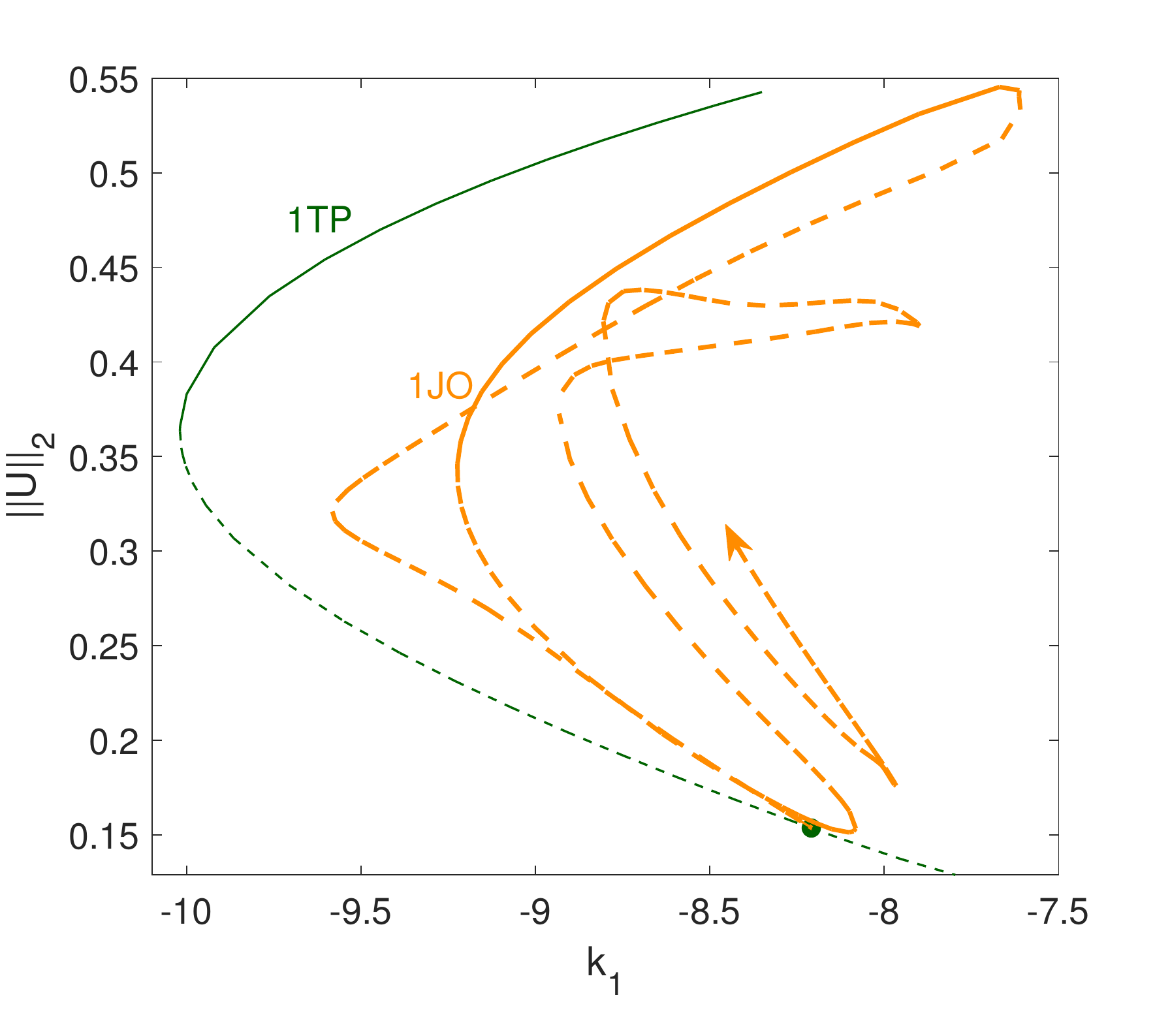}
	\caption{(a) A second branch of modulated TW states (brown) that emerges from the second Hopf point 
		(left brown bullet) on the 1TP branch (green); the branch of 1JO (gray) is shown in the background. Parameters are as in Fig.~2 of the main text. (b) Partial branch of 1TP (green) and a bifurcating (green bullet) branch of 1JO (orange) at $D_v=20$. Other parameters are as in (a). In both figures, solid lines indicate stable solutions.}
	\label{fig:mTW}
\end{figure}

In the main text (Fig.~2), we showed the 1JO solutions that bifurcate from the first Hopf bifurcation of the traveling pulses on the 1TP branch. In Fig.~\ref{fig:mTW}(a), we show an additional example of the emergence of mTW (brown) that bifurcate from a subsequent Hopf bifurcation (brown bullet). The branch reconnects back to the 1TP branch via another Hopf bifurcation at a lower $||U||_2$ (also marked by brown bullet). This behavior is typical for many of the mTW branches that bifurcate from the 1TP branch.

When parameters are changed, Hopf points may be created or destroyed. For instance, decreasing $D_v$ from $D_v=25$ we find that the first Hopf point, responsible for our primary 1JO branch, annihilates with the second Hopf point, resulting in a shift in the onset of the 1JO branch. Figure~\ref{fig:mTW}(b) shows that at $D_v=20$ the 1JO branch bifurcates at $k_1\approx -8.2$. At first, the bifurcating mTP branch follows the former brown branch [see (a)] towards more negative $k_1$ before following the branch obtained at $D_v=25$, and so needs an extra fold before the solutions turn into genuine (stable) JOs.
\pagebreak

\section{Jumping target waves}\label{SM:JW}

\begin{figure}[h]
	\centering
	(a)\includegraphics[width=0.95\textwidth]{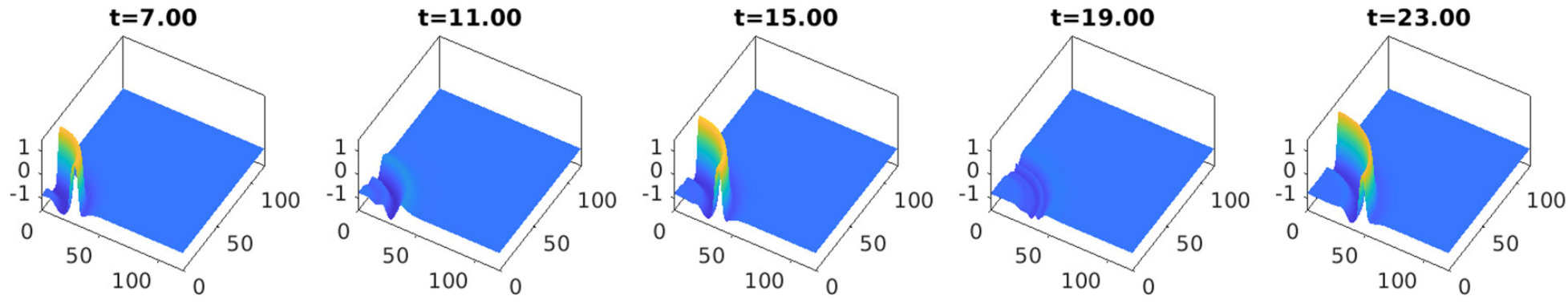}
	(b)\includegraphics[width=0.95\textwidth]{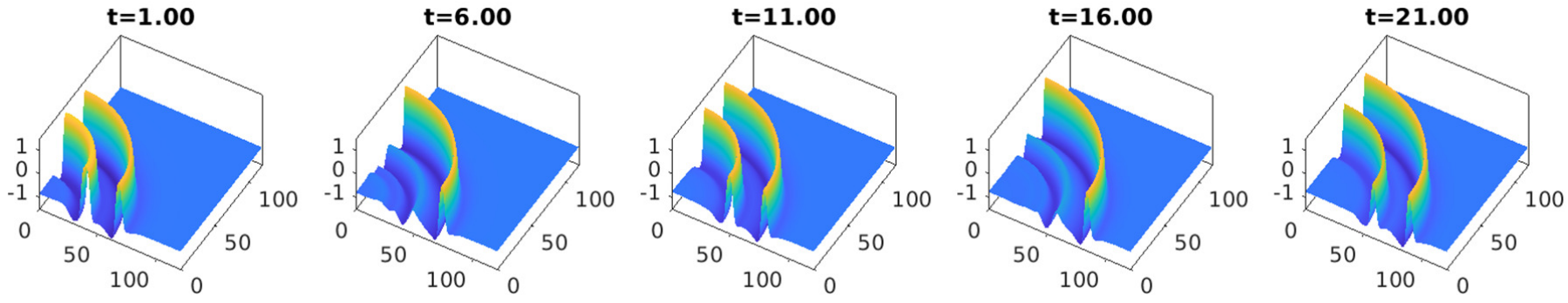}
	\caption{{DNS showing the evolution of jumping waves (JWs) $u(x,y,t)$ at selected times over two temporal periods. (a) 1JW and (b) 1TP1JW, corresponding, respectively, to Figs.~1(a) and 1(d) of the main text. See also the supplementary movies m1 and m2. In both cases the initial condition corresponds to an axisymmetric version of the corresponding 1D solution and boundary conditions are of Neumann (no-flux) type.}}
	\label{fig:JW}
\end{figure}

In the main text, we have presented a study of JOs in 1D and here in Fig.~\ref{fig:JW} and supplementary movies m1 and m2 we demonstrate, using DNS, that solutions such as those displayed in Figs.~1(a,d) can also be obtained in 2D. Solutions similar to those in Fig.~\ref{fig:JW}(a) have been found in experiments by~\citet{cherkashin2008discontinuously}. 
	
\end{document}

%% file: hudef.tex
\def\bexe{\begin{exercise}}\def\eexe{\eex\end{exercise}}
\def\bsol{\begin{solution}}\def\esol{\eex\end{solution}}
\def\bexa{\begin{example}}\def\eexa{\end{example}}
\def\brem{\begin{remark}}\def\erem{\end{remark}}
\def\bthm{\begin{theorem}}\def\ethm{\end{theorem}}
\def\blem{\begin{lemma}}\def\elem{\end{lemma}}
\def\bcor{\begin{corollary}}\def\ecor{\end{corollary}}
\def\bdefi{\begin{definition}}\def\edefi{\end{definition}}
\newcommand{\IDEA}{\textbf{Idea of the Proof.} }

\def\bmip{\begin{minipage}{\textwidth}}\def\emip{\end{minipage}}

\newcommand{\R}{{\mathbb R}}

\def\vt{\vartheta}\def\pa{{\partial}}
\newcommand{\bi}{\begin{itemize}}\newcommand{\ei}{\end{itemize}}
\newcommand{\ben}{\begin{enumerate}}\newcommand{\een}{\end{enumerate}}
\newcommand{\bce}{\begin{center}}\newcommand{\ece}{\end{center}}
\newcommand{\bci}{\begin{compactitem}}\newcommand{\eci}{\end{compactitem}}
\newcommand{\bcen}{\begin{compactenum}}\newcommand{\ecen}{\end{compactenum}}
\newcommand{\bcena}{\begin{compactenum}[(a)]}
\newcommand{\reff}[1]{(\ref{#1})}

\newcommand{\spr}[1]{\left\langle #1 \right\rangle}

\newcommand{\barr}{\begin{array}}\newcommand{\earr}{\end{array}}
\newcommand{\bpm}{\begin{pmatrix}}\newcommand{\epm}{\end{pmatrix}}
\newcommand{\bsm}{\left(\begin{smallmatrix}}
\newcommand{\esm}{\end{smallmatrix}\right)}
\newcommand{\ba}{\begin{array}}\newcommand{\ea}{\end{array}}
\def\dd{\, {\rm d}}

\def\re{{\rm Re\,}}\def\im{{\rm Im\,}}

\def\ddt{\frac{\rm d}{{\rm d}t}}
\def\del{\delta}

\def\eex{\hfill\mbox{$\rfloor$}}

\def\sig{\sigma}

\def\bd{\begin{displaymath}} \def\ed{\end{displaymath}}
\def\ba{\begin{array}} \def\ea{\end{array}}
